\documentclass[11 pt,twoside]{article}
\usepackage{amsmath,amsfonts,amssymb}
\usepackage{latexsym}
\usepackage{dcolumn}
\usepackage{graphicx,epsfig}
\usepackage{amsthm}
\usepackage{color}
\usepackage{framed}

\begin{document}

\setcounter{page}{1}

\pagestyle{plain} \vspace{1 cm}

\begin{center}
{\large{\bf {Geodesic Structure of the Quantum-Corrected Schwarzschild Black Hole Surrounded by Quintessence}}}\\

\vspace{1 cm}
{\bf Kourosh Nozari$^{\dag,}$}\footnote{knozari@umz.ac.ir (Corresponding Author)}\quad and \quad {\bf Milad Hajebrahimi$^{\dag,}$}\footnote{m.hajebrahimi@stu.umz.ac.ir}\\

\vspace{0.5 cm}
$^{\dag}$Department of Theoretical Physics, Faculty of Basic Sciences, University of Mazandaran,\\
P. O. Box 47416-95447, Babolsar, Iran\\

\end{center}

\vspace{1 cm}

\begin{abstract}
By considering the back-reaction of the spacetime through the spherically symmetric quantum fluctuations of the background metric, Kazakov and Solodukhin removed the singularity of the Schwarzschild black hole. This regular Schwarzschild black hole has a spherical central region with a radius of the order of the Planck length. On the other hand, due to the positively accelerating expansion of the Universe, it seems that there exists a universal repulsive force known as dark energy. In the framework of quantum field theories, the quintessence field is a candidate model for investigating and modeling dark energy. Accordingly, by taking into account the quintessential matter field in the background of the Schwarzschild black hole, Kiselev gained the metric of this black hole surrounded by quintessence. By combining these two above ideas, in the present study we consider the quantum-corrected Schwarzschild black hole surrounded by quintessence to investigate null and time-like geodesics structure. Generally, this study points out that black holes are quantum-gravitational objects. We will show that the accelerated expansion of the Universe, instead of dark energy, happens because of the presence of quantum effects in this setup. Also, due to the presence of the central Planck-size sphere, the regular black hole has been possessed a shifting over radial coordinate in its inner structure. 
\end{abstract}

\newpage
\section{Introduction}

In both frameworks of physics and mathematics, there are some terrible situations known as singularity, where theories and mathematical setups can not apply to them. In these situations, a quantity or a variable goes to mathematical infinity and becomes unexplainable. In physics, one can see several such situations, for example, the singularity of the electric field of a point-like particle at its position. Also, the general theory of relativity proposed by Albert Einstein in 1915, has two famous singularities in its outcomes: the singularity at the center of black holes, and the Big Bang singularity. The black hole singularity is a central region in the inner structure of a black hole, where the curvature of the spacetime becomes infinite. Rotating black holes have a ring-shape singularity region, while non-rotating ones have a point-like singularity region. Although such a singularity has arisen from the general theory of relativity, this theory cannot describe or even remove it.

A lot of studies focused on performing some novel ideas to remove the black hole singularity regions, theoretically to gain regular black holes \cite{Modesto2004, Dymnikova1992, Bojowald2005, Modesto2011, Gambini2008, Trodden1993, Olmo2015, Beato1999, Cadoni1995, Shankaranarayanan2004, Nashed2002, Bambi2013, Beato1998, Bronnikov2001, Balart2014, Berej2006, Junior2015, Myung2009, Matyjasek2008, Uchikata2012, Rodrigues2016, Lorenzo2016, Nicolini2005, Olmo2012}, and the possibility of forming them \cite{Zhang2015, Hayward2006, Mbonye2005, Hossenfelder2010, Giugno2018}. Accordingly, by taking into account these regular black holes, some studies have focused on investigating the features and behaviors of them \cite{Maluf2018, Abdujabbarov2017, Easson2003, Abdujabbarov2016, Flachi2013, ZLi2014, Myung2007, ZLi2013, Toshmatov2015, JLi2013, Jusufi2018, Tharanath2015, Balart2010}. On the other hand, quantum gravity outcomes have led to some extraordinary impacts on the fundamentals of physics \cite{Abdo2009, Das2008, Basilakos2010, Hamber1995, Alfaro2002, Taveras2008, Nozari2005a, Nozari2007a}, specifically in the field of thermodynamics \cite{Nozari2006a, Nozari2019, Kamali2016, Nozari2009a, Nozari2008a, Tawfik2013}. Moreover, it is well known that black holes have lots of quantum aspects \cite{Harvey1992, Carlip1995, Mann1992, Giddings1994, Majumdar2000}. Studies in this area indicate that quantum, and also, quantum gravity effects lead to some important corrections in the spirit of the black hole physics \cite{Nozari2006b, Nozari2005b, Nozari2013a, Nozari2013b, Nozari2012a, Nozari2010, Nozari2009b, Nozari2008b, Nozari2007b, Reuter2011, Modesto2008, Soltani2014, Nozari2015, Nozari2012b, Nozari2005c, Nozari2008c, Saghafi2017, Solodukhin1995, Meissner2004, Nouicer2007, Nozari2006c, Nozari2007c, Nozari2008d}. Therefore, it seems that one of the most applicable ways to remove black hole singularity regions is to bring the quantum effects, particularly the outcomes of quantum field theories into the framework of black hole physics \cite{Abedi2016}.

In search of a quantized gravity to remove the singularity region of the Schwarzschild (SCH) black hole, Kazakov and Solodukhin in 1994 admitted that the background metric and matter fields have spherically symmetric quantum excitations \cite{Kazakov1994}. Also, they considered the back-reaction of spacetime \cite{Abedi2016, Lousto1988}, due to such quantum fluctuations. These considerations led to a quantum-corrected Schwarzschild (QS) black hole, theoretically. Such a black hole has the metric coefficient to the form of $$f_{_{00}}(r)=\frac{1}{f_{_{11}}(r)}=-\frac{2M}{r}+\frac{1}{r}\sqrt{r^{2}-a^{2}}$$ in which the new parameter, $a$ can be seen. This parameter is a minimal distance with the length dimension, as expected and, it is of the order of the Planck length, $l_{pl}$. As Kazakov and Solodukhin noticed, this result is universal. Therefore, it appears that such a decision is independent of whether background matter is involved or not \cite{Kazakov1994}. The singularity of the QS black hole, which was originally at $r=0$, is shifted to $$r=a\equiv r_{min}\sim l_{pl}\,,$$ which means that it is now a two-dimensional sphere of radius $r=a$. Recently, the Hawking radiation via tunneling process for the QS black hole has been studied in Ref. \cite{Hajebrahimi2020}.

According to the recent astronomical and astrophysical data, for example, observations of type Ia supernova \cite{Perlmutter1999, Riess1998, Garnavich1998, Riess1999, Lapuente1995, Riess2004}, cosmic microwave background (CMB) data \cite{Spergel2007, Miller1999, Hanany2000, Lamon2006}, and large scale structure (LSS) results \cite{Tegmark2004, Seljak2005}, it can be said that the Universe is in an accelerating era with positive acceleration. An accelerating Universe indicates that there is a kind of gravity, which is repulsive. Repulsive gravity is not identified, exactly. It is like the energy of the increasing rate of the expansion of the Universe. Therefore, it has been termed as dark energy. So, it seems that we are in the dark energy dominant epoch of the evolution of the Universe. Nobody knows what is dark energy, exactly. But, there are several cosmological models to describe it, in which all of them indicate that dark energy has negative pressure to accelerate the Universe. The Cosmological Constant is a cosmological model for describing dark energy. This simple model gives a very small observed value for dark energy, which leads to the problem of fine-tuning \cite{Weinberg1989}. However, there exist some alternative models for dark energy, based on scalar fields \cite{Copeland2006} including quintessence \cite{Ratra1988, Carroll1998, Caldwell1998, Zlatev1999, Sahni2000, Matos2001, Capozziello2006}, chameleon fields \cite{Khoury2004}, $\kappa$-essence \cite{Picon2000, Scherrer2004, Chimento2005}, tachyon field \cite{Padmanabhan2002}, phantom dark energy \cite{Caldwell2002}, dilaton dark energy \cite{Gasperini2002}, and so on.

The quintessence, like other alternative models, is a scalar field coupled to gravity, which is dynamic and spatially inhomogeneous (instead of Cosmological Constant). Since the quintessence does not form structures (despite matter), it should be very light, which leads to a large Compton wavelength for it. In 2003, Kiselev derived a spacetime metric for the Schwarzschild black hole in the background of quintessential matter to the form of $$h_{_{00}}(r)=\frac{1}{h_{_{11}}(r)}=1-\frac{2M}{r}-\frac{c}{r^{3\omega_{q}+1}}\,$$ in which $c$ is a positive normalization factor associated with quintessence field, and $\omega_{q}$ is the parameter of the corresponding equation of state. In order to solve gravitational field equations, Kiselev used the quintessence stress-energy tensor, so that, it had the conditions of additivity and linearity as $T_{tt}=T_{rr}=\rho_{q}$ and $T_{\theta \theta}=T_{\phi \phi}=-\big(\rho_{q}\big(3\omega_{q}+1\big)/2\big)$ \cite{Kiselev2003}. As this new derivation published, several studies on the Schwarzschild black hole surrounded by quintessence (SQ) have been released, including the studying of null geodesics structure for SQ black hole \cite{Fernando2012}, the studying of time-like geodesics structure for SQ black hole \cite{Uniyal2015}, thermodynamics of Schwarzschild and Reissner–Nordstr\"{o}m black holes surrounded by quintessence \cite{Ghaderi2016}, and the studying of null geodesics for Schwarzschild-anti de Sitter black hole surrounded by quintessence field \cite{Malakolkalami2015}.

By taking into account these two above ideas of Kazakov-Solodukhin and Kiselev, one can find the metric of the quantum-corrected (non-singular) Schwarzschild black hole surrounded by quintessence (QSQ). In 2019, Shahjalal, considered such a deformed black hole to analyze its thermodynamics \cite{Shahjalal2019}. Also, the effects of quantum corrections on the criticality and efficiency of black holes surrounded by a perfect fluid have been studied in Ref. \cite{Bezerra2019}.

Studying geodesic structure around black holes is an interesting issue in the literature \cite{Sharp1979, Stuchl1991, Amarilla2010, Cruz2005, Wilkins1972, Cruz1994, Bret2002, Leiva2009, Abbas2014, Stuchl2015, Sheng2011}. In this study, we take into account the QSQ black hole to analyze the structures of corresponding null and time-like geodesics, respectively. We also compare the results of the present study with the outcomes of the SQ black hole in Refs. \cite{Fernando2012, Uniyal2015}, and also, the geodesic structure of the SCH and QS black holes.

First, in Section 2, we introduce the features of this regular black hole, and then, we find the corresponding Hawking temperature. In Section 3, we analyze the null geodesics structure of the black hole for radial geodesics and geodesics with angular momentum. Then, we find the bending angle for such a black hole. In Section 4, we study the time-like geodesic structure. Finally, in Section 5, we draw to an end with some important conclusions.

\section{Some features of QSQ black hole}

The line element of the QSQ black hole is as follows \cite{Shahjalal2019}
\begin{equation}
ds^{2}=-g(r)dt^{2}+\frac{1}{g(r)}dr^{2}+r^{2}\left(d\theta^{2}+\sin^{2}\theta\,d\phi^{2}\right)\,,
\end{equation}
where $g(r)$ is obtained as follows
\begin{equation}
g(r)=-\frac{2M}{r}+\frac{1}{r}\sqrt{r^{2}-a^{2}}-\frac{c}{r^{3\omega_{q}+1}}\,,
\end{equation}
where $M$ is the black hole mass ($G=c^{2}=1$), $c$ is a normalization factor depend on the distribution function of the energy density of the quintessence field, and $a$ is the quantum correction factor with the dimensionality of length as mentioned in Introduction. This modified black hole has a central spherical region with the radius $r=a$ of the order of the Planck length, instead of the point-like singularity of the SCH black hole. By computing the Ricci Scalar for this regular black hole, one can prove that, like the SCH solution, it is asymptotically flat. The quintessential matter field surrounding this regular black hole has an equation of state of the form $P_{q}=\omega_{q}\rho_{q}$ in which the energy density, $\rho_{q}$ of the field is as below \cite{Kiselev2003}
\begin{equation}
\rho_{q}=-\frac{c}{2}\frac{3\omega_{q}}{r^{3\omega_{q}+3}}\,,
\end{equation}
where $\omega_{q}$ is the equation of state parameter, which ranges in $-1\leq\omega_{q}\leq-\frac{1}{3}$ \cite{Uniyal2015}. While the pressure, $P_{q}$ for the quintessential matter is negative, the energy density, $\rho_{q}$ is always positive. Now, we choose $\omega_{q}=-\frac{2}{3}$, and accordingly, we rewrite Eq. (2) for the metric coefficient as follows \cite{Shahjalal2019}
\begin{equation}
g(r)=-\frac{2M}{r}+\frac{1}{r}\sqrt{r^{2}-a^{2}}-cr\,.
\end{equation}
The choice of $\omega_{q}=-\frac{2}{3}$ is a special case, which is chosen just for simplifying the following calculations and having explicit solutions \cite{Ghaderi2016, Malakolkalami2015}.

As $8Mc<1-4a^{2}c^{2}$, the above metric coefficient has two roots associated with two horizons of the non-singular hole as below
\begin{equation}
r_{inner}=\frac{\sqrt{1-4Mc-\sqrt{1-8Mc-4a^{2}c^{2}}}}{\sqrt{2}\,c}\,,
\end{equation}
\begin{equation}
r_{outer}=\frac{\sqrt{1-4Mc+\sqrt{1-8Mc-4a^{2}c^{2}}}}{\sqrt{2}\,c}\,.
\end{equation}
Between the two horizons, i.e., at $r_{inner}<r<r_{outer}$, the $t$ coordinate is time-like, and the $r$ coordinate is space-like. But at the region $a<r<r_{inner}$, the $t$ coordinate is space-like, and the $r$ coordinate is time-like. Therefore, the inner horizon, $r_{inner}$ is like the event horizon of the SCH black hole. The outer horizon, $r_{outer}$ known as quintessence horizon is a cosmological horizon, like the cosmological horizon of the de Sitter-Schwarzschild spacetime. With respect to Eq. (6) for the outer horizon, which is the cosmological horizon, i.e., $R\approx 1.3\times 10^{10}$, and through numerical analysis, one can find an approximate value for the quantum correction term as $a\approx 10^{-9}$ for which the values of $M=10^{19}$ and $c=0.11$ is considered. This approximate value once again has been deduced in Section 3. At $r_{inner}<r<r_{outer}$, the spacetime is static, where is the region between the two horizons. So, we focus on the motion of photons and particles in this region. Hence, the spacetime for the QSQ black hole is similar to a regular de Sitter-Schwarzschild spacetime. But for $8Mc=1-4a^{2}c^{2}$, the regular black hole has a degenerate horizon, and for $8Mc>1-4a^{2}c^{2}$, it has a naked, spherical central region with radius $r=a$.

\begin{figure}
  \centering
  \includegraphics[width=0.7\textwidth]{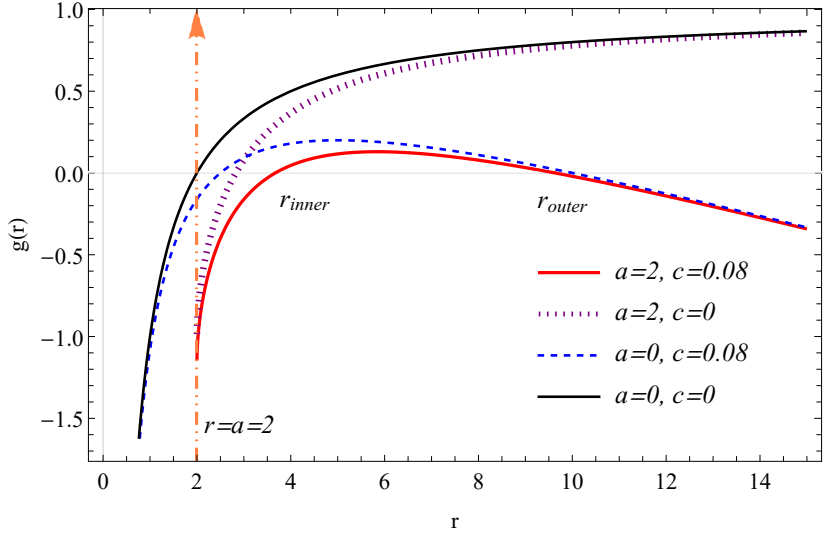}
  \caption{\label{Fig1}\small{\emph{The plot of $g(r)$ versus $r$ for QSQ, QS, SQ, and SCH cases in which we set $M=1$.}}}
\end{figure}
Figure 1 shows the function of the metric coefficient (4) versus $r$ for four cases of QSQ, QS, SQ, and SCH. From Figure 1, we see a cutoff at $r=a$ for the curves of QSQ and QS. We set $a=2$, but this choice is just for clarity and simplicity in illustrating the figure and all following figures. The cutoff line of both cases QSQ and QS, and the two horizons of the case QSQ have been displayed in Figure 1. The quintessence horizon of both QSQ and SQ cases are matchable together, approximately. Also, the SQ curve approaches the QSQ curve, and QS curve approaches SCH curve, asymptotically. The curve of the metric coefficient for the SCH case goes to infinity by increasing $r$. Consequently, it appears that the role of the quintessence term is to produce the cosmological horizon, and to limit the values of the metric coefficient, and the role of the quantum-correction term is to cut off the metric coefficient curve across the two-dimensional central region, which leads to shift the inner structure of black hole due to the presence of Planck-size central sphere.

One can calculate the Hawking temperature for the QSQ black hole by using the following equation
\begin{equation}
T_{r_{inner},r_{outer}}=\frac{1}{4\pi}\left|\frac{dg(r)}{dr}\right|_{r=r_{inner},r_{outer}}\,.
\end{equation}
Eq. (7) results in two following relations for Hawking temperature associated with the inner horizon and the outer one, respectively
\begin{equation}
T_{r_{inner}}=\frac{c\left(1-8Mc-\sqrt{1-8Mc-4a^{2}c^{2}}-\frac{2\sqrt{2}a^{2}c^{2}}{\sqrt{\left(1-4Mc-
2a^{2}c^{2}-\sqrt{1-8Mc-4a^{2}c^{2}}\right)}}\right)}{4\pi\left(-1+4Mc+\sqrt{1-8Mc-4a^{2}c^{2}}\right)}\,,
\end{equation}
\begin{equation}
T_{r_{outer}}=\frac{c\left(1-8Mc+\sqrt{1-8Mc-4a^{2}c^{2}}-\frac{2\sqrt{2}a^{2}c^{2}}{\sqrt{\left(1-4Mc-
2a^{2}c^{2}+\sqrt{1-8Mc-4a^{2}c^{2}}\right)}}\right)}{4\pi\left(1-4Mc+\sqrt{1-8Mc-4a^{2}c^{2}}\right)}\,.
\end{equation}

\begin{figure}
  \centering
  \includegraphics[width=0.7\textwidth]{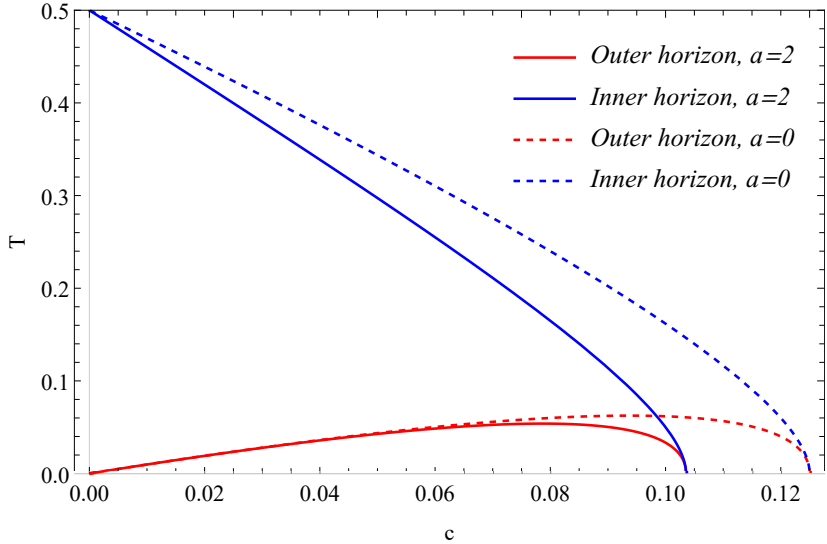}
  \caption{\label{Fig2}\small{\emph{The Hawking temperature for both inner and outer horizons with respect to $c$ for which we set $M=1$.}}}
\end{figure}
Figure 2 displays the Hawking temperature for QSQ and SQ cases associated with both inner and outer horizons, so that the temperature of the inner horizon is bigger than the outer one, and also, by comparison, both of them are smaller than the corresponding results of the SQ case in Ref. \cite{Fernando2012}.

Henceforward, we consider the QSQ black hole to attain its geodesic structure and compare the achieved results with SQ and SCH cases. The known geodesic equations and the Lagrangian, i.e., its constraint equations have the following forms
\begin{eqnarray}
\ddot{x}^{\mu}+\Gamma_{\nu\lambda}^{\mu}\dot{x}^{\nu}\dot{x}^{\lambda}=0\,, \\
2K=g_{\mu\nu}\dot{x}^{\mu}\dot{x}^{\nu}=e\,,
\end{eqnarray}
where `dot' denotes to differentiate with respect to the proper time, $s$. In the last equation, $e$ is either $0$ or $-1$: the case with $e=0$ is for massless particles traveling on the null geodesics with the speed of light and the case with $e=-1$ is for massive particles traveling on the time-like geodesics. In the next two sections, we first consider the massless case and then the massive case to drive the null geodesics and the time-like ones, respectively for the black hole described in Eqs. (1) and (4).

\section{Null Geodesics Structure}

In this section, we focus on null geodesics associated with massless particles. Such particles travel with the speed of light. So, they move on the edge of the light cone. For the null geodesics, one can find out Eq. (11) with respect to Eqs. (1) and (4) as follows
\begin{equation}
2K=-\left(-g(r)\left(\frac{dt}{ds}\right)^{2}+\frac{1}{g(r)}\left(\frac{dr}{ds}\right)^{2}+r^{2}\left(\frac{d\theta}{ds}
\right)^{2}+r^{2}\sin^{2}(\theta)\left(\frac{d\phi}{ds}\right)^{2}\right)=0\,.
\end{equation}

The QSQ black hole has two Killing vectors, $\partial_t$ and $\partial_\phi$. Consequently, there are two constants of motion as follows
\begin{eqnarray}
g(r)\dot{t}\equiv E_{n}\,,\\
r^{2}\sin^{2}(\theta)\dot{\phi}\equiv L\,.
\end{eqnarray}

We suppose that the motion is on the equatorial plane. So, we set $\theta=\frac{\pi}{2}$, and hence, we conclude $\dot{\theta}=\ddot{\theta}=0$. Now, the Lagrangian (12) with respect to Eqs. (13) and (14) takes the following form
\begin{equation}
E_{n}^{2}=\dot{r}^{2}+g(r)\left(\frac{L^{2}}{r^{2}}\right)\,.
\end{equation}
We define the effective potential, $V_{e}$ as follows
\begin{equation}
V_{e}\equiv g(r)\left(\frac{L^{2}}{r^{2}}\right)\,,
\end{equation}
and then, we can rewrite Eq. (18) in a simple form as the following relation
\begin{equation}
E_{n}^{2}=\dot{r}^{2}+V_{e}\,.
\end{equation}
By dividing Eq. (17) to Eq. (20), one can gather a differential equation between $\phi$ and $r$ as below
\begin{equation}
\frac{d\phi}{dr}=\frac{L}{r^{2}\sqrt{E_{n}^{2}-V_{e}}}\,.
\end{equation}
We follow this procedure for two cases: first for the radial null geodesics and second for the null geodesics with angular momentum ($L\neq 0$).

\subsection{Radial Null Geodesics}

A radial null geodesic is used to describe the motion of a massless test particle, which travels from some region out of the black hole and then approaches its center, directly ($\phi=0$). Such a motion has no angular momentum. So, the effective potential in Eq. (19) becomes zero ($V_{e}=0$) and the equations of motion for $\dot{t}$ and $\dot{r}$ in Eqs. (13) and (17) takes the following forms, respectively
\begin{eqnarray}
\frac{dt}{ds}=\frac{E_{n}}{g(r)}\,,\\
\frac{dr}{ds}=\pm E_{n}\,,
\end{eqnarray}

From Eqs. (19) and (20), one can find out the following differential equation between $t$ and $r$
\begin{equation}
\frac{dt}{dr}=\pm\frac{1}{g(r)}=\pm\left(-\frac{2M}{r}+\frac{1}{r}\sqrt{r^{2}-a^{2}}-cr\right)^{-1}\,.
\end{equation}
This equation simply leads to an exact relation between $t$ and $r$ as below
\begin{equation}
t=\pm\Bigg\{\frac{-1}{c\sqrt{\big(8Mc+4a^{2}c^{2}-1\big)}}\arctan\bigg[\frac{2c\sqrt{(r^{2}-a^{2})}-1}
{{\sqrt{\big(8Mc+4a^{2}c^{2}-1\big)}}}\bigg]+\frac{1}{2c}\ln\Big[\sqrt{r^{2}-a^{2}}-2M-cr^{2}\Big]\Bigg\}+C_{\pm}.
\end{equation}
In this equation, the integration constants, $C_{\pm}$ are imaginary, which result in some real values for $t$. Also, the minus sign is for outgoing motion, and the plus sign is for ingoing motion. In addition to $t$, we can also value the proper time, $s$, by integrating Eq. (20) as following
\begin{equation}
s=\pm\frac{r}{E_{n}}+C_{\pm}.
\end{equation}

\begin{figure}
  \centering
  \includegraphics[width=0.7\textwidth]{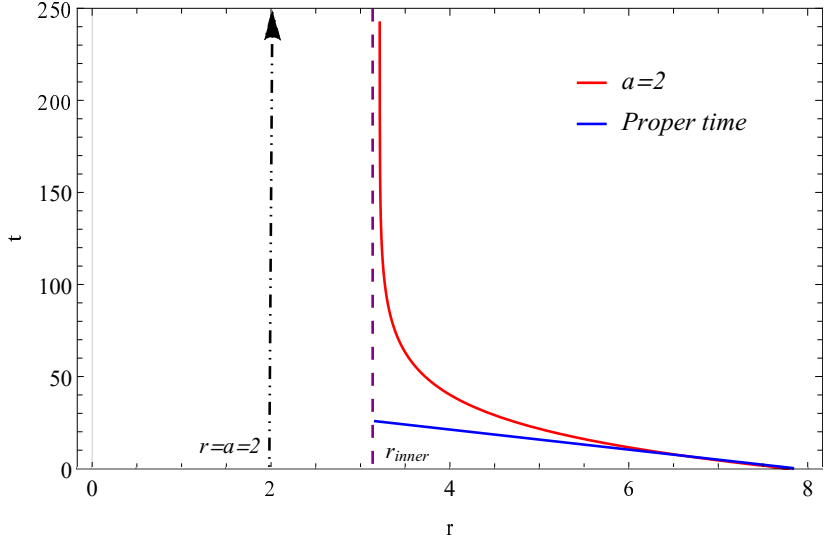}
  \caption{\label{Fig3}\small{\emph{The curves of the $t$ and $s$ versus $r$ for QSQ case in which we set $M=1$, $c=0.05$.}}}
\end{figure}
\begin{figure}
  \centering
  \includegraphics[width=0.7\textwidth]{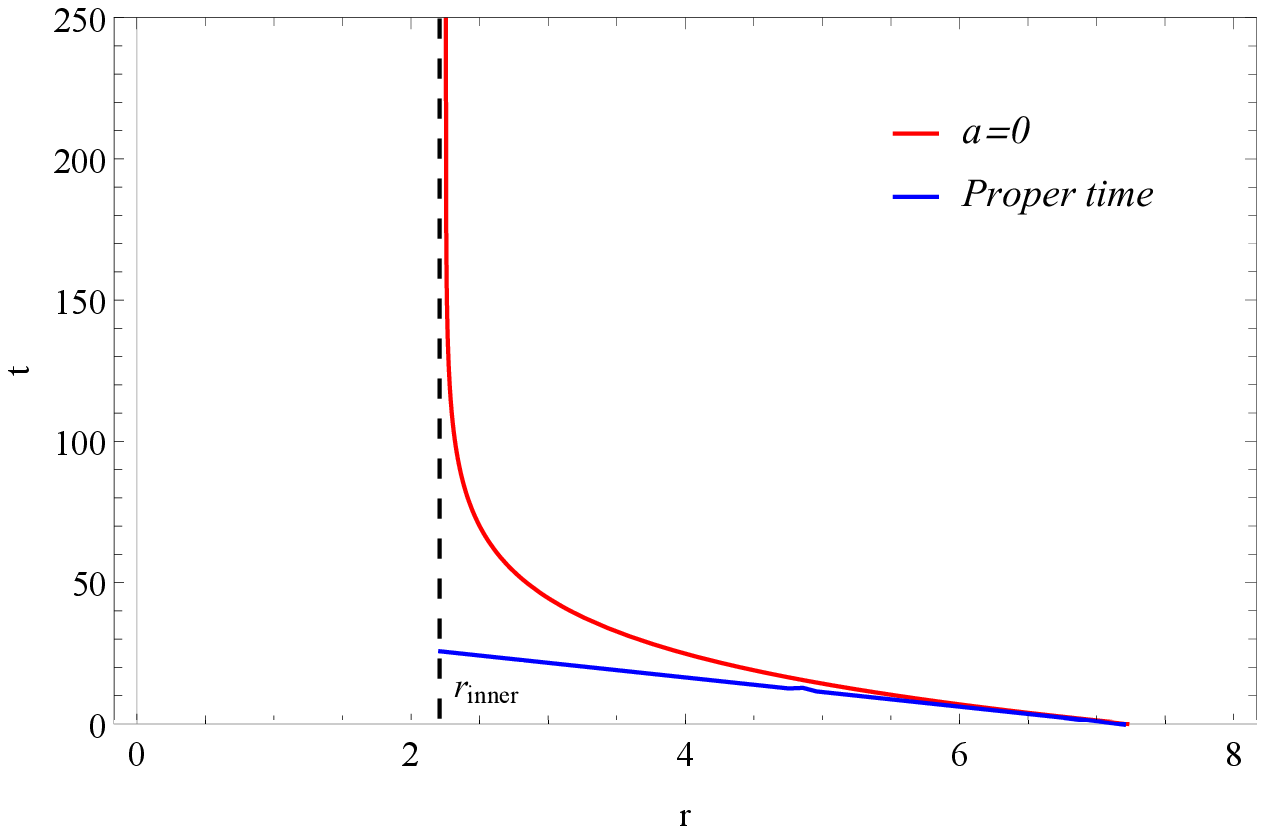}
  \caption{\label{Fig4}\small{\emph{The curves of $t$ and $s$ versus $r$ for SQ case in which we set $M=1$, $c=0.05$.}}}
\end{figure}
\begin{figure}
  \centering
  \includegraphics[width=0.7\textwidth]{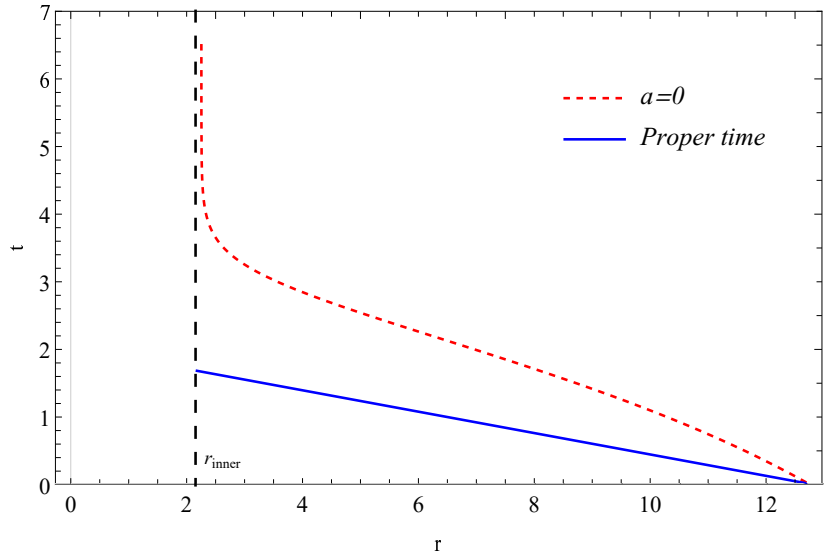}
  \caption{\label{Fig5}\small{\emph{The curves of $t$ and $s$ versus $r$ for SQ case from the Ref. \cite{Fernando2012}.}}}
\end{figure}
Figure 3 depicts the curves of both $t$ and $s$ in the QSQ case for ingoing motion. It seems that due to the quantum effects, the inner structure of the black hole shifted to the value of $r=a$. Also, for comparison, Figure 4 is the curve of $t$ and $s$ in the SQ case. Due to the absence of quantum effects, Figure 4 does not have any shifting in its inner structure. So, it seems that quantum effects have a regulatory role in the black hole physics, which modify the inner structure of the black hole. We note that the illustrated figure for $t$ for SQ case in Ref. \cite{Fernando2012}, which is seen in Figure 5 has not been correctly drawn up.

\subsection{Null Geodesics with Angular Momentum ($L\neq 0$)}

Null geodesics with angular momentum instead of the radial ones can rotate or circle the black holes in some particular orbits. Studying such geodesics is an interesting area in black hole physics.

\subsubsection{Effective Potential}

According to Eq. (16) with $L\neq 0$, one can read the effective potential in QSQ black hole as follows
\begin{equation}
V_{e}\equiv g(r)\bigg(\frac{L^{2}}{r^{2}}\bigg)=-\frac{2ML^{2}}{r^{3}}+\frac{L^{2}}{r^{3}}\sqrt{r^{2}-a^{2}}-\frac{cL^{2}}{r}\,.
\end{equation}
Figure 6 shows the plot of $V_{e}$ versus $r$ for different values of $c$ in four cases of QSQ, QS, SQ, and SCH. In this figure, we see a cutoff at $r=a=2$ for the curves of QSQ and QS. Since $r_{inner}$ and $r_{outer}$ are the two roots of the function of $g(r)$, the effective potential in the of QSQ, like SQ case, becomes zero at the inner horizon and the outer one, as it is seen from Figure 6. But in two cases of SCH and QS, the function of the effective potential has only one root at the Schwarzschild radius, $r_{_{SCH}}=2M$, and the quantum-corrected Schwarzschild radius, $r_{_{QS}}=\sqrt{4 M^{2}+a^{2}}$, respectively. Also, the QSQ curve approaches the SQ curve, and the QS curve approaches SCH curve, asymptotically. From Figure 6, it seems that the role of the quintessence term is to reduce the values of the effective potential, and to make two roots for the effective potential associated with two horizons. However, the role of the quantum deformation term in addition to reduce the values of the effective potential is to shift the potential curve to somewhere bigger than $r=a$. Nevertheless, the values of this potential in the SCH case are bigger than QSQ, QS, and SQ cases. Therefore, it seems that quantum effects and quintessence term have a regulatory role in black hole physics.

\begin{figure}
  \centering
  \includegraphics[width=0.7\textwidth]{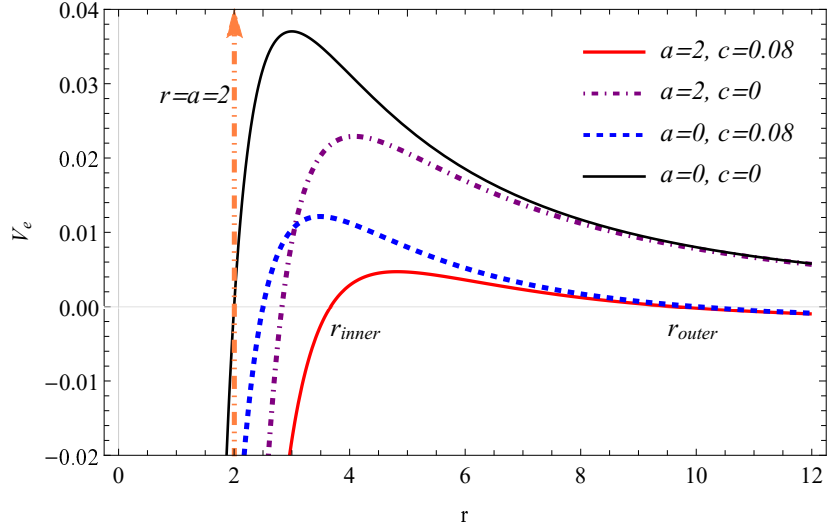}
  \caption{\label{Fig6}\small{\emph{The plot of $V_{e}$ versus $r$ for QSQ, QS, SQ, and SCH cases in which we set $M=1$, $c=0.08$, $a=2$, and $L=1$.}}}
\end{figure}

\begin{figure}
  \centering
  \includegraphics[width=0.7\textwidth]{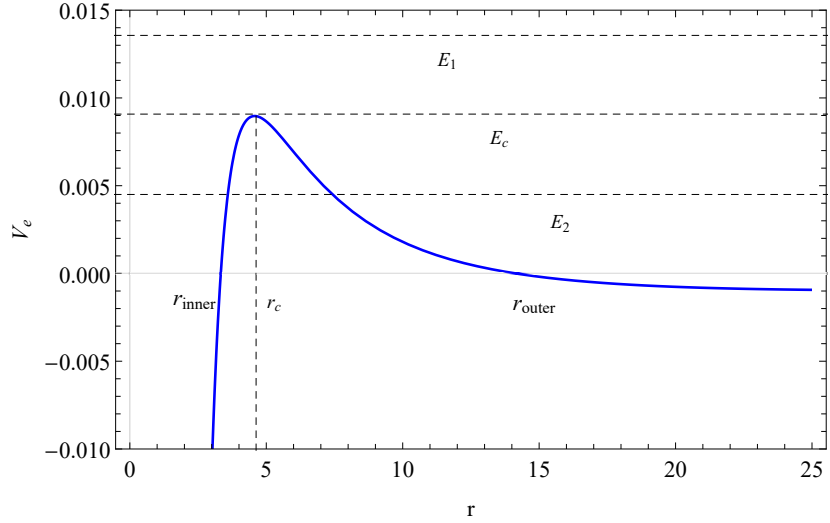}
  \caption{\label{Fig7}\small{\emph{The relation between the energy levels, $E_{n}$, and the effective potential for QSQ case in which we set $M=1$, $a=2$, $c=0.06$, and $L=1$.}}}
\end{figure}
According to Eq. (17), which presents the relation between the energy $E_{n}$ and the effective potential $V_{e}$ for QSQ case, one can deduce that these energy levels govern the motion of particles. In Figure 7, one can see three levels $E_{1}$, $E_{c}$, and $E_{2}$ so that $E_{n}=E_{c}$ leads to $E_{n}^{2}-V_{e}=\dot{r}^{2}=0$ for which the orbits are circular. It is obvious from Figure 7 that the circular orbits are unstable at $r=r_{c}$. The situations $E_{n}=E_{1}$, and $E_{n}=E_{2}$ leads to $E_{n}^{2}-V_{e}=\dot{r}^{2}\geq 0$ so that for the case $E_{1}$, due to the quantum corrections term, photons fall over the (modified) spherical region with radius $r=a$.

\subsubsection{Circular Orbits}

As mentioned above, in the case of $E_{n}=E_{c}$ for QSQ black hole, we have unstable, null circular orbits for which $V_{e}=E_{c}^{2}$ and so, the radial gradient of $V_{e}$ is zero, i.e.,
\begin{equation}
\frac{dV_{e}}{dr}=0\,.
\end{equation}
Eq. (25) with respect to Eq. (24) leads to six solutions for $r$ as follows
\begin{equation}
r_{_{1,2}}=\pm\left(\frac{1}{c}\right)\sqrt{\frac{1}{3A_{1}}\left[2^{^{\frac{1}{3}}}A_{2}+A_{1}\left(2^{^{-\frac{1}{3}}}A_{1}+A_{3}\right)\right]}\,,
\end{equation}
\begin{equation}
r_{_{3,4}}=\pm\left(\frac{1}{c}\right)\sqrt{\frac{1}{3A_{1}}\left[(-1)^{^{\frac{2}{3}}}2^{^{\frac{1}{3}}}A_{2}+A_{1}\left(-2^{^{-\frac{4}{3}}}\left(1+i\sqrt{3}\right)
A_{1}+A_{3}\right)\right]}\,,
\end{equation}
\begin{equation}
r_{_{5,6}}=\pm\left(\frac{1}{c}\right)\sqrt{\frac{1}{3A_{1}}\left[(-2)^{^{\frac{1}{3}}}(-A_{2})+A_{1}\left(2^{^{-\frac{4}{3}}}\left(-1+i\sqrt{3}\right)A_{1}
+A_{3}\right)\right]}\,,
\end{equation}
where $A_{1}$, $A_{2}$, and $A_{3}$ are defined as below
\begin{eqnarray}
A_{1}\equiv\Bigg\{128+a^{2}c^{2}\bigg(2a^{4}c^{4}+3a^{2}c^{2}\Big(53+12Mc\Big)+24\Big(3Mc\big(16+3Mc\big)-14\Big)\bigg)\nonumber\\+144Mc\Big(3Mc\big(5+Mc\big)
-8\Big)+9c^{2}\Big(a^{2}c+4M\Big)\sqrt{3}\bigg[a^{4}c^{2}\Big(4a^{2}c^{2}+72Mc+59\Big)\nonumber\\+144M^{2}\Big(6Mc-1\Big)+8a^{2}\Big(3Mc\big(7+18Mc\big)
-4\Big)\bigg]^{^{\frac{1}{2}}}\Bigg\}^{^{\frac{1}{3}}}\,,
\end{eqnarray}
\begin{equation}
A_{2}\equiv a^{2}c^{2}\left(a^{2}c^{2}+12Mc\right)-4c^{2}\left(7a^{2}-9M^{2}\right)+16\left(1-6Mc\right)\,,
\end{equation}
\begin{equation}
A_{3}\equiv 4\left(1-3Mc\right)+a^{2}c^{2}\,.
\end{equation}
Three of these solutions are nonphysical because of their minus sign. So, we only take into account three solutions with a plus sign. By numerical analysis, one can determine that $r_{_{2}}$ and $r_{_{6}}$ are bigger and smaller than the outer and the inner event horizons, respectively. Since we are focusing on the motion of photons within two horizons, we take into account $r_{_{4}}$ as the radius of unstable, null circular orbits, and we call it $r_{c}$ \cite{Hod2011, Pradhan2011}. The unstable, null circular orbits specify the black hole by an unstable region \cite{Pugliese2011}. From Eq. (27), we see that the unstable, null circular orbit for QSQ black hole is as follows
\begin{eqnarray}
r_{c}=\left(\frac{1}{c}\right)\sqrt{\frac{1}{3A_{1}}\left[(-1)^{^{\frac{2}{3}}}2^{^{\frac{1}{3}}}A_{2}+A_{1}\left(-2^{^{-\frac{4}{3}}}\left(1+i\sqrt{3}\right)
A_{1}+A_{3}\right)\right]}\,,
\end{eqnarray}
while for the SCH black hole, this unstable, null circular orbit is at $r=3M$.

\begin{figure}
  \centering
  \includegraphics[width=0.7\textwidth]{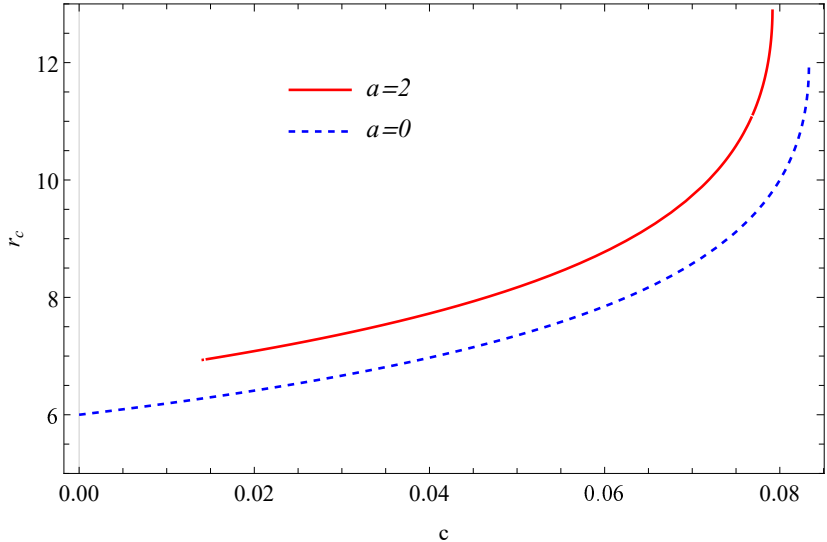}
  \caption{\label{Fig8}\small{\emph{The illustration of $r_{c}$ versus $c$ for QSQ and SQ cases in which we set $M=2$.}}}
\end{figure}
In Figure 8, the curve of $r_{c}$ versus $c$ has been illustrated for both QSQ and SQ cases. The curve $r_{c}$ in the QSQ case is bigger than the SQ case in Ref. \cite{Fernando2012}. It seems that the bigger the value of $c$, the bigger the value of $r_{c}$, as mentioned in Ref. \cite{Fernando2012}. In addition, it seems that in the QSQ case, one could not find a value for $r_{c}$ for small values of $c$.

As it is seen from Figure 7, the effective potential $V_{e}$ at $r=r_{c}$ has an unstable nature. This fact leads to a remarkable outcome. The unstable, null circular orbits with radius $r_{c}$ (where the potential is not stable) create a null hypersurface at $r=r_{c}$ around the black hole known as the photon sphere, which is a remarkable subject in black hole physics \cite{Claudel2001}. On this sphere with radius $r=r_{c}$, all of the massless particles including photons circle the black hole on the unstable, null circular geodesics \cite{Decanini2010}. From a mathematical point of view, the photon sphere is the locus of all unstable, null circular orbits with radius $r_{c}$ circling the black hole.

By using Eq. (17) for null circular orbits ($\dot{r}=0$) and Eq. (24), one can write an equation between energy level, $E_{c}$ and the angular momentum, $L_{c}$ of these unstable, null circular orbits for QSQ black hole as follows
\begin{equation}
\left(\frac{E_{c}}{L_{c}}\right)^{2}=\frac{g(r_{c})}{r_{c}^{2}}=\frac{\left(-2M+\sqrt{r_{c}^{2}-a^{2}}-cr_{c}^{2}\right)}{r_{c}^{3}}=
\left(\frac{1}{D_{c}}\right)^{2}\,,
\end{equation}
where $D_{e}$ interprets the impact parameter (the ratio of angular momentum to linear momentum) of these unstable, null circular orbits. From Eq. (33), one can find $D_{c}$ in terms of $c$ and $a$ as follows
\begin{equation}
D_{c}=\sqrt{\frac{r_{c}^{3}}{\left(-2M+\sqrt{r_{c}^{2}-a^{2}}-cr_{c}^{2}\right)}}\,.
\end{equation}

\begin{figure}
  \centering
  \includegraphics[width=0.7\textwidth]{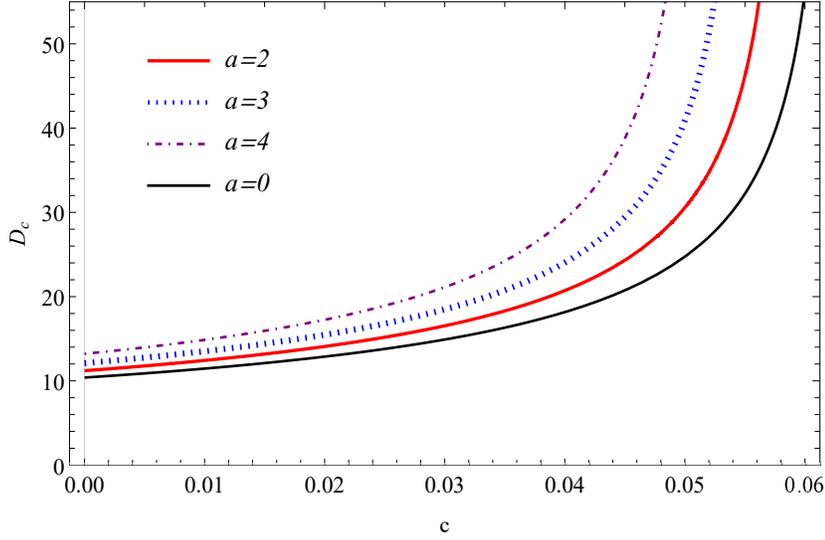}
  \caption{\label{Fig9}\small{\emph{The illustration of $D_{c}$ versus $c$ for both QSQ case with different values of $a$, and SQ case with $a=0$ in which we set $M=2$.}}}
\end{figure}
\begin{figure}
  \centering
  \includegraphics[width=0.7\textwidth]{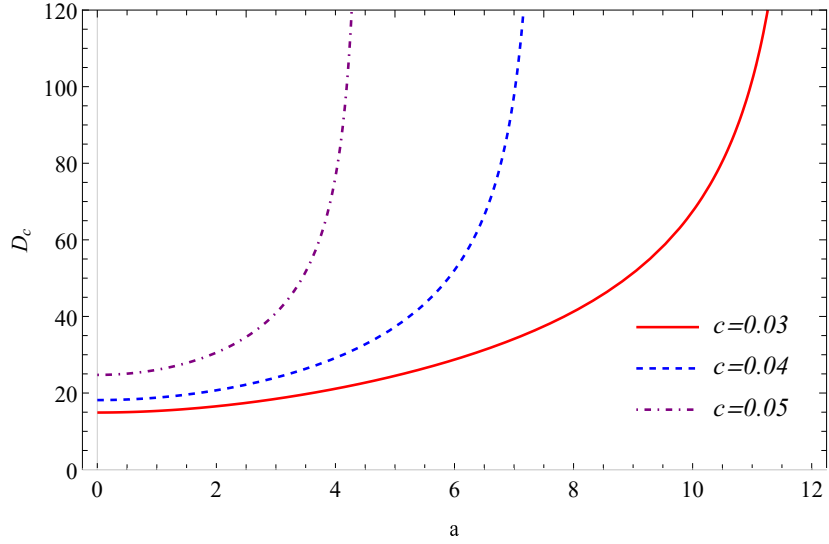}
  \caption{\label{Fig10}\small{\emph{The plot of $D_{c}$ versus $a$ for QSQ case with different values of $c$ in which we set $M=2$.}}}
\end{figure}
\begin{figure}
  \centering
  \includegraphics[width=0.7\textwidth]{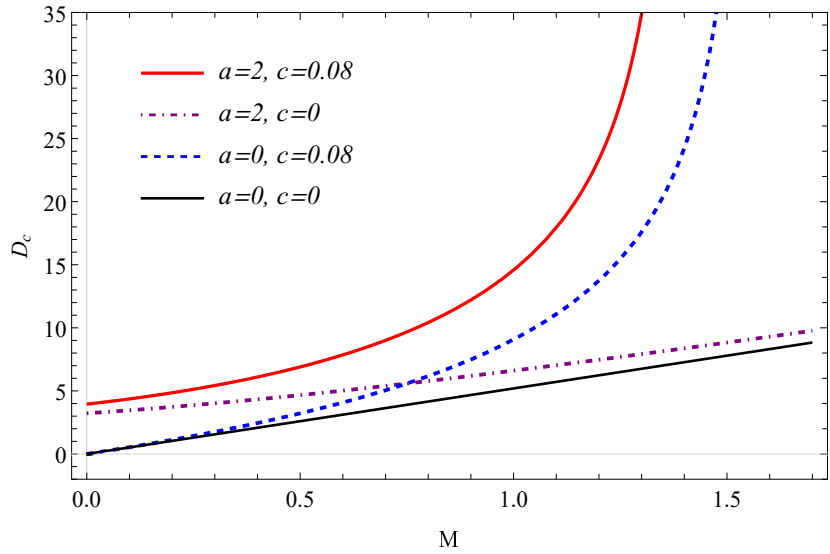}
  \caption{\label{Fig11}\small{\emph{The illustration of $D_{c}$ versus $M$ for three cases of QSQ, QS, SQ, and SCH.}}}
\end{figure}
Figure 9 shows the illustration of $D_{c}$ in terms of $c$ with different values of $a$ for QSQ and SQ cases. It seems that as $c$ values are increasing, $D_{c}$ values become to increase. Also, Figure 9 indicates that for larger values of $a$, the larger values of $D_{c}$ is obtainable. From Figure 9, we see that the curve with $a=0$ associated with SQ case has smaller values than the curves of the QSQ case with $a\neq 0$ for all values of $c$. So, the role of quantum effects is to amplify the values of $D_{c}$. On the other hand, Figure 10 shows the illustration of $D_{c}$ versus $a$ with different values of $c$ for the QSQ case. It depicts that increasing in values of $a$ leads to an increase in values of $D_{c}$. Also, Figure 10 indicates that for larger values of $c$, the much larger values of $D_{c}$ is obtainable. Figure 11 is the curve of $D_{c}$ in terms of $M$ for the QSQ, QS, SQ, and SCH cases. From Figure 11, it seems that the curve of $D_{c}$ in the QSQ case is bigger than two other cases without the quantum-correction term. Also, the QS curve approaches the SCH curve, asymptotically. As it is seen from Figure 11, the curves of QSQ, and QS cases begin from some non-zero $D_{c}$.

\subsubsection{Period of the Unstable, Null Circular Orbits}

From Eq. (14) for $r_{c}$ and $L_{c}$, one can find the following relation
\begin{equation}
\int_{0}^{T_{s}}ds=\int_{0}^{2\pi}\left(\frac{r_{c}^{2}}{L_{c}}\right)d\phi\,,
\end{equation}
where $T_{s}$ is the period in terms of the proper time $s$. This relation leads to the following result for $T_{s}$ in the QSQ black hole
\begin{equation}
T_{s}=\frac{2\pi r_{c}^{2}}{L_{c}}\,.
\end{equation}
Since $r_{c}$ for the SCH black hole is $3M$, the corresponding period $T_{s}$ in terms of the proper time $s$ is as below
\begin{equation}
T_{s,SCH}=\frac{18\pi M^{2}}{L_{c}}\,.
\end{equation}

One can combine Eqs. (13) and (33) to find the period in terms of the time coordinate $t$ for the QSQ case as follows
\begin{equation}
T_{t}=\frac{2\pi r_{c}}{\sqrt{g_{_{qq}}(r_{c})}}\,.
\end{equation}
For the SCH black hole, one can deduce the following result for the period in terms of the time coordinate $t$
\begin{equation}
T_{t,SCH}=6\pi\sqrt{3}M\,.
\end{equation}

\begin{figure}
  \centering
  \includegraphics[width=0.7\textwidth]{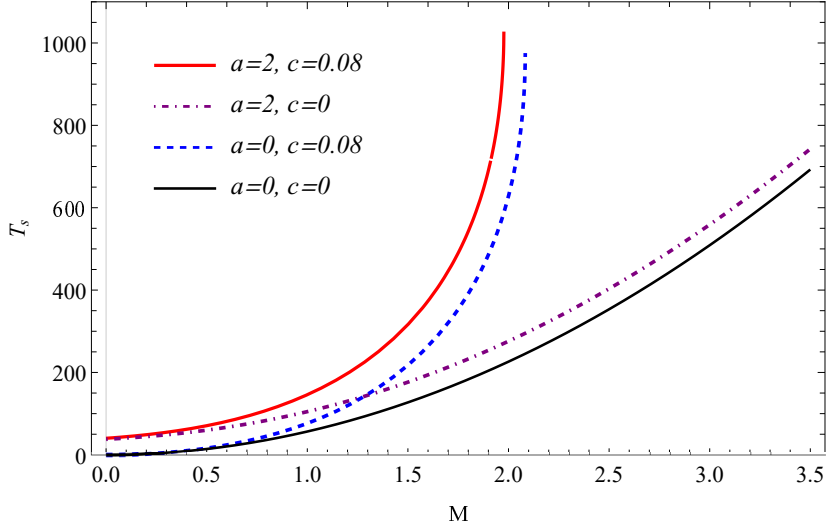}
  \caption{\label{Fig12}\small{\emph{The illustration of $T_{s}$ versus $M$ for QSQ, QS, SQ, and SCH cases in which we set $L_{c}=1$.}}}
\end{figure}
The plot of $T_{s}$ with respect to $M$ in Figure 12 shows that the curve of the QSQ case has some bigger values than the curve of the SQ case, and also, both of them do not go to infinite values. But the SCH case infinitely increases by the increasing of $M$. Also, the QS curve approaches SCH curve, asymptotically. The curves of QSQ, and QS cases begin from some non-zero $T_{s}$. From Figure 12, one can see that the bigger the size of the black hole, the bigger the period of the circular, null orbits in terms of the proper time, $s$.

\begin{figure}
  \centering
  \includegraphics[width=0.7\textwidth]{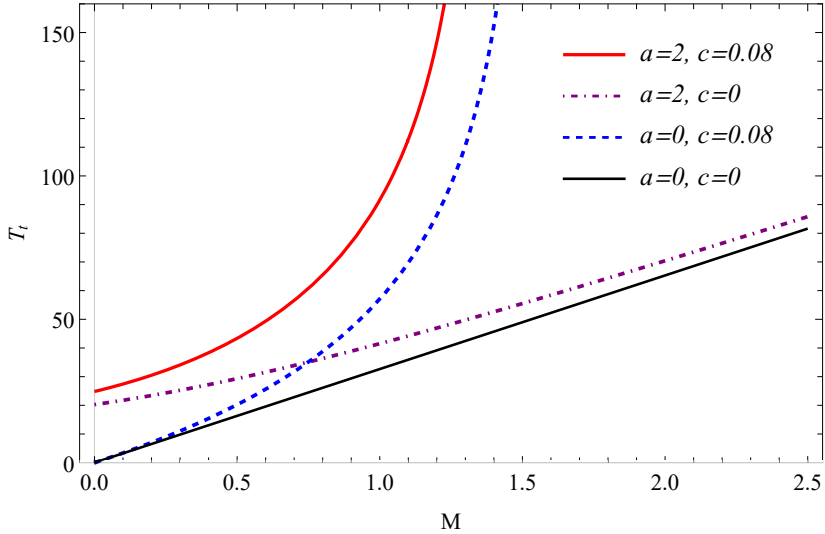}
  \caption{\label{Fig13}\small{\emph{The illustration of $T_{t}$ versus $M$ for QSQ, QS, SQ, and SCH cases.}}}
\end{figure}
Figure 13 depicts the curve of $T_{t}$ versus $M$ for QSQ, QS, SQ, and SCH cases. While SCH and QS cases infinitely increase as a straight line by increasing $M$, the QSQ case approaches the SQ case, asymptotically. As it is seen from Figure 13, the curves of QSQ, and QS cases begin from some non-zero $T_{t}$. In summary, the bigger the size of the black hole, the bigger the period of the circular, null orbits in terms of the time coordinate, $t$.

\subsubsection{Lyapunov Exponent for the Unstable, Null Circular Orbits}

In mathematics, a quantity, $\lambda$ known as the Lyapunov exponent, characterizes the rate of separation of infinitesimally close trajectories in a dynamical system. In other words, this quantity measures an average rate for convergence ($\lambda<0$) and divergence ($\lambda>0$) of nearby trajectories in the phase space. In Ref. \cite{Cardoso2009}, an expression for $\lambda$ for a black hole is computed as follows \cite{Fernando2012}
\begin{equation}
\lambda=\sqrt{-\frac{V''_{e}\left(r_{c}\right)}{2\left(\dot{t}\left(r_{c}\right)\right)^{2}}}=\sqrt{-\frac{V''_{e}\left(r_{c}\right)r_{c}^{2}g\left(r_{c}\right)}{2L_{c}}}\,.
\end{equation}

In Figure 14, the curve of $\lambda$ versus $c$ shows that the values of $\lambda$, and accordingly, the separation of unstable, null circular orbits in the QSQ case are smaller than the SQ ones. Also, it seems that in the QSQ case, $\lambda$ cannot obtain for small values of $c$. Due to the positivity of the values of $\lambda$ in both scenarios, the unstable, null circular orbits are divergence as we expected.
\begin{figure}
  \centering
  \includegraphics[width=0.7\textwidth]{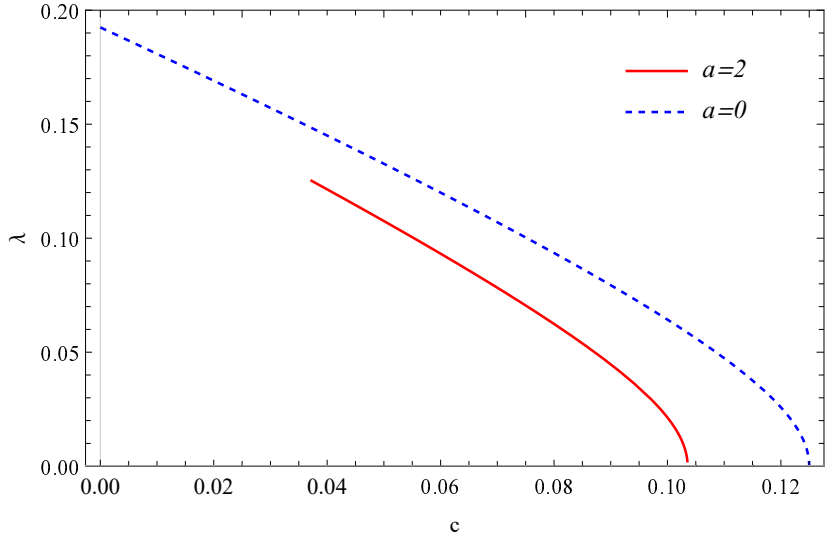}
  \caption{\label{Fig14}\small{\emph{The illustration of $\lambda$ versus $c$ for QSQ and SQ cases in which we set $L_{c}=1$ and $M=1$.}}}
\end{figure}

\subsubsection{Force on Massless Particles, Like Photons}

By using the idea of the gradient of effective potentials to lead to effective forces, one can find the effective force on massless particles like photons around QSQ black hole as follows
\begin{equation}
F_{e}=-\frac{1}{2}\frac{dV_{e}}{dr}=-\frac{3L^{2}M}{r^{4}}-\frac{cL^{2}}{2r^{2}}+\frac{2L^{2}r^{2}-3a^{2}L^{2}}{2r^{4}\sqrt{r^{2}-a^{2}}}\,.
\end{equation}
From the equation of motion (17), we have added a factor of $\frac{1}{2}$ in Eq. (41) \cite{Fernando2012}.

In Eq. (41), the sign of the first term is negative, so it is attractive and such a term is known as the Newtonian term. Similarly, the second term sign is negative, which again means that this term is attractive. On the other hand, the second term is associated with the quintessential matter. Since the quintessential matter is a candidate for dark energy, in this scheme, the dark energy ingredient of the Universe is attractive, while it should be a repulsive force to boost acceleration, generally! Since we choose a static system between two horizons, such a result is expected \cite{Fernando2012}. However, the last term in Eq. (41) has a positive sign. So this term is repulsive. Since this term is the quantum-corrected term, it seems that such a term provides a repulsive force instead of the dark energy component to facilitate the acceleration of the Universe. This is because of the presence of back-reaction in the setup \cite{Buchert2015}. Consequently, in this study, it seems that the quintessential matter plays an attractive role to decelerate the expansion of the Universe, and the quantum effects make the Universe more accelerated.

\begin{figure}
  \centering
  \includegraphics[width=0.7\textwidth]{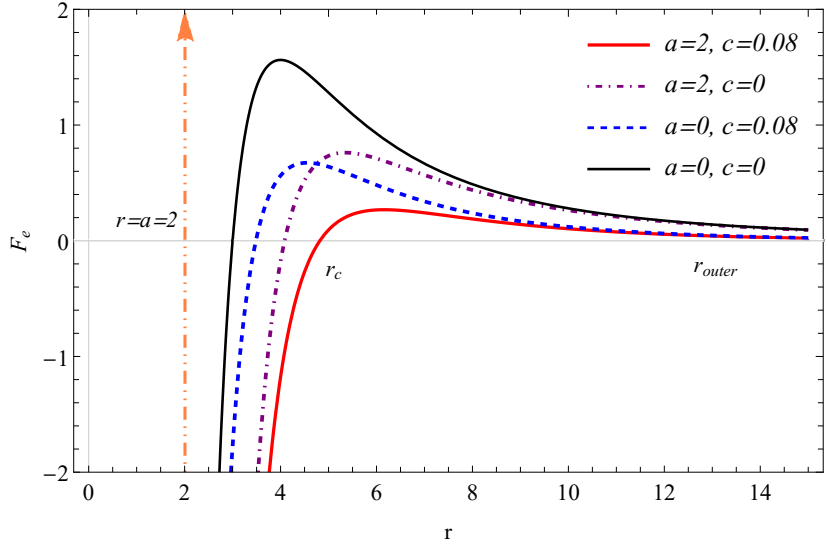}
  \caption{\label{Fig15}\small{\emph{The illustration of $F_{e}$ versus $r$ for QSQ, QS, SQ, and SCH cases for positive values of $F_{e}$ in which we set $L=20$ and $M=1$.}}}
\end{figure}
\begin{figure}
  \centering
  \includegraphics[width=0.7\textwidth]{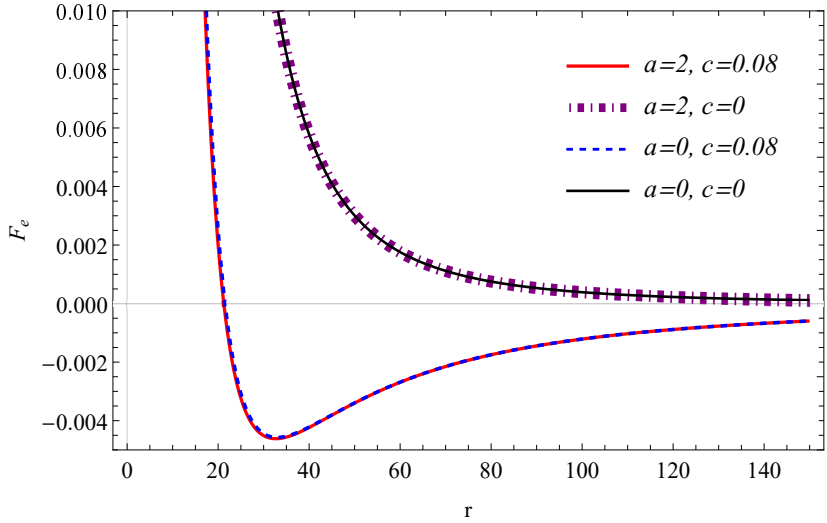}
  \caption{\label{Fig16}\small{\emph{The illustration of $F_{e}$ versus $r$ for QSQ, QS, SQ, and SCH cases for negative values of $F_{e}$ in which we set $L=20$ and $M=1$.}}}
\end{figure}
Figure 15 and Figure 16 are the plots of $F_{e}$ versus $r$ in two ranges of $r$ for QSQ, QS, SQ, and SCH cases. The values of $F_{e}$ in the QSQ case are much smaller than the SQ case in the range of $r_{inner}<r<r_{outer}$. Moreover, in all of these cases, at the point $r=r_{c}$, the result of $F_{e}=0$ can be obtained. Also, for $r_{inner}<r<r_{c}$ the effective force is negative and for $r_{c}<r<r_{outer}$ the force is positive. For $r>r_{outer}$, the effective force in the QSQ case has a behavior, as same as the SQ case, approximately. The curve of the effective force for two cases of SCH and QS in Figure 15 and Figure 16 shows us that the effective force is positive in the corresponding range $r>r_{c}$. On the other hand, due to quantum effects in this scheme, the curves of the QSQ, and QS cases are shifted to the $r=a=2$. Consequently, the role of the quantum effects is to shift the function of the effective force to somewhere bigger than $r=a$, and also, is to reduce the amount of the force in its positive ranges. Since in this scheme, the quantum effects are responsible to accelerate the Universe as mentioned above, it seems that quantum effects also lead to a cutoff on this acceleration by reducing the amount of the effective force. While the role of the quintessential matter is to make cosmological horizon, and to reduce the amount of effective force, a little. However, for large amounts of $r$ the effective force tents to zero in all of these cases as expected from the asymptomatically flatness of the spacetime at far from the black hole.

\subsection{Analysis of Null Geodesics by Using the Variable $u=\frac{1}{r}$}

In the study of geodesic structure, it is well known that the change of variable $u=\frac{1}{r}$ is more useful for analyzing the geodesics equations of motion. So, by such a change of variable, Eq. (18) with respect to Eq. (16) becomes
\begin{equation}
\left(\frac{du}{d\phi}\right)^{2}=\Psi(u)\,,
\end{equation}
where we have
\begin{equation}
\Psi(u)=2Mu^{3}-u^{2}\sqrt{1-u^{2}a^{2}}+cu+\frac{E_{n}^{2}}{L^{2}}\,.
\end{equation}

Figure 17 presents the behavior of $\Psi(u)$ in Eq. (43) versus $u$ for QSQ, QS, SQ, and SCH cases. In the used parameter space and by numerical analysis, the QSQ and QS cases, and also, the SQ and SCH cases are matched together, respectively. In this figure, it seems that both $\Psi(u)$ and $u$ can not go to $\pm\infty$ in two cases of QSQ and QS. As it is seen, at the point $u=\pm\frac{1}{a}$, where we set $a=2$, the curve of $\Psi(u)$ is limited. This result is due to the presence of quantum effects in this scheme. Also, like the SQ case, $u=0$ leads to $\Psi(u)=\frac{E_{n}^{2}}{L^{2}}$. Moreover, the SCH and the QSQ cases have the same behavior, approximately.
\begin{figure}
  \centering
  \includegraphics[width=0.7\textwidth]{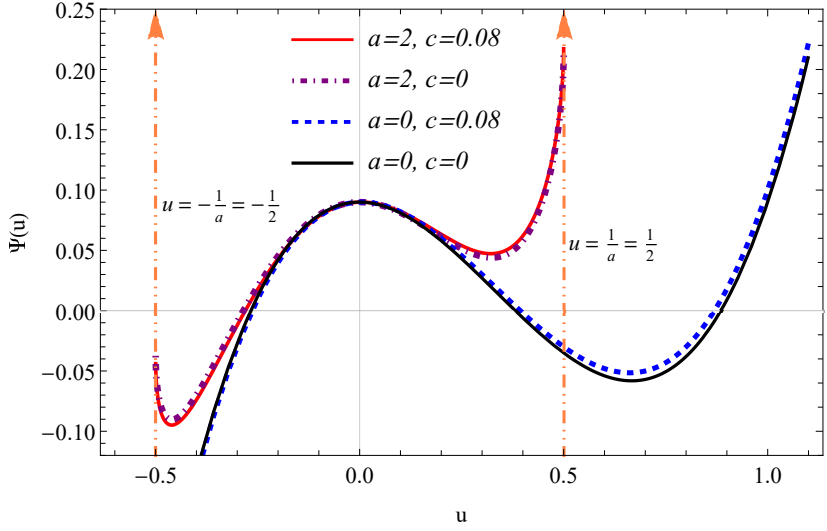}
  \caption{\label{Fig17}\small{\emph{The illustration of $\Psi(u)$ versus $u$ for QSQ, QS, SQ and SCH cases in which we set $L=30$, $E_{e}=9$ and $M=0.5$.}}}
\end{figure}

To solve Eq. (42), it is required to find the roots of Eq. (43), i.e., it needs to solve $\Psi(u)=0$. But, the roots of this equation have not explicit form. Hence, we expand Eq. (43) to eliminate higher-order terms of $a$, which leads to the following result
\begin{equation}
\Psi(u)\approx 2Mu^{3}-u^{2}+\frac{a^{2}u^{4}}{2}+cu+\frac{E_{n}^{2}}{L^{2}}\,.
\end{equation}

Now, it is easy to find the roots of Eq. (44) as are listed below
\begin{equation}
u_{1}=-\frac{M}{a^{2}}-\sqrt{B_{2}}-\sqrt{B_{3}+\frac{a^{4}c+2a^{2}M+4M^{3}}{2a^{6}\sqrt{B_{2}}}}\,,
\end{equation}
\begin{equation}
u_{2}=-\frac{M}{a^{2}}-\sqrt{B_{2}}+\sqrt{B_{3}+\frac{a^{4}c+2a^{2}M+4M^{3}}{2a^{6}\sqrt{B_{2}}}}\,,
\end{equation}
\begin{equation}
u_{3}=-\frac{M}{a^{2}}+\sqrt{B_{2}}-\sqrt{B_{3}-\frac{a^{4}c+2a^{2}M+4M^{3}}{2a^{6}\sqrt{B_{2}}}}\,,
\end{equation}
\begin{equation}
u_{4}=-\frac{M}{a^{2}}+\sqrt{B_{2}}+\sqrt{B_{3}-\frac{a^{4}c+2a^{2}M+4M^{3}}{2a^{6}\sqrt{B_{2}}}}\,,
\end{equation}
where we have defined
\begin{eqnarray}
B_{1}\equiv 3a^{2}\Bigg\{\bigg[\Big(288a^{2}\big(E_{n}/L\big)^{2}+108a^{2}c^{2}+864\big(E_{n}/L\big)^{2}M^{2}+144cM-16\Big)^{2}\nonumber\hspace{-0.5 cm}\\-4\Big(24a^{2}\big(E_{n}/L\big)^{2}-24cM+4\Big)^{3}\bigg]^{\frac{1}{2}}\nonumber\hspace{2 cm}\\
+288a^{2}\big(E_{n}/L\big)^{2}+108a^{2}c^{2}+864\big(E_{n}/L\big)^{2}M^{2}+144cM-16\Bigg\}^{\frac{1}{3}}\,,
\end{eqnarray}
\begin{equation}
B_{2}\equiv\frac{1}{3a^{2}}+\frac{M^{2}}{a^{4}}+\frac{B_{1}}{36a^{4}\sqrt[3]{2}}+\frac{2^{^{\frac{1}{3}}}\left(6a^{2}\left(E_{n}/L\right)^{2}
-6cM+1\right)}{B_{1}}\,,
\end{equation}
\begin{equation}
B_{3}\equiv\frac{2}{3a^{2}}+\frac{2M^{2}}{a^{4}}-\frac{B_{1}}{36a^{4}\sqrt[3]{2}}-\frac{2^{^{\frac{1}{3}}}\left(6a^{2}\left(E_{n}/L\right)^{2}
-6cM+1\right)}{B_{1}}\,.
\end{equation}
By using these roots, one can rewrite Eq. (44) as below
\begin{equation}
\Psi(u)=\left(u-u_{1}\right)\left(u-u_{2}\right)\left(u-u_{3}\right)\left(u-u_{4}\right)\,.
\end{equation}
Moreover, between these roots, one can check the following relation
\begin{equation}
u_{1}+u_{2}+u_{3}+u_{4}=-\frac{4M}{a^{2}}\,.
\end{equation}

Now, by using Eq. (52), one can rewrite Eq. (42) as the following differential equation
\begin{equation}
\frac{du}{\sqrt{\Psi(u)}}=\frac{du}{\sqrt{\left(u-u_{1}\right)\left(u-u_{2}\right)\left(u-u_{3}\right)\left(u-u_{4}\right)}}=d\phi\,.
\end{equation}
This easily can be integrated to lead to the following relation
\begin{equation}
\phi=2\frac{\mathrm{EllipticF}[\mu,z]}{\sqrt{\left(u_{1}-u_{3}\right)\left(u_{2}-u_{4}\right)}}+\mathrm{constant}\,,
\end{equation}
where we have defined
\begin{equation}
\sin\mu=\frac{\sqrt{u-u_{1}}\sqrt{u_{2}-u_{4}}}{\sqrt{u-u_{2}}\sqrt{u_{1}-u_{4}}}\,,
\end{equation}
\begin{equation}
z=\frac{\left(u_{1}-u_{4}\right)\left(u_{2}-u_{3}\right)}{\left(u_{1}-u_{3}\right)\left(u_{2}-u_{4}\right)}\,,
\end{equation}
and $\mathrm{EllipticF}[\mu,z]$ is the Elliptic Integral of the first kind. So, Eq. (55) gives the trajectory of $u-\phi$ for QSQ black hole.

\subsubsection{Circular Orbits and Unbounded Orbits}

Due to the nature of $\Psi(u)$ versus $u$, the function of $\Psi(u)$ has degenerate roots for the case of the circular orbits so that, Eq. (52) becomes as the following equation
\begin{equation}
\Psi(u)=\left(u-u_{1}\right)\left(u-u_{2}\right)\left(u-u_{c}\right)^{2}\,.
\end{equation}
Generally, $u_{c}$ is equal to the inversion of $r_{c}$, i.e., $u_{c}=\frac{1}{r_{c}}$. But, in this case, since we expand $\Psi(u)$ around $a=0$, such equality is not holed. Figure 18 is a qualitative plot of $\Psi(u)$ in Eq. (58) versus $u$ in the QSQ case.
\begin{figure}
  \centering
  \includegraphics[width=0.7\textwidth]{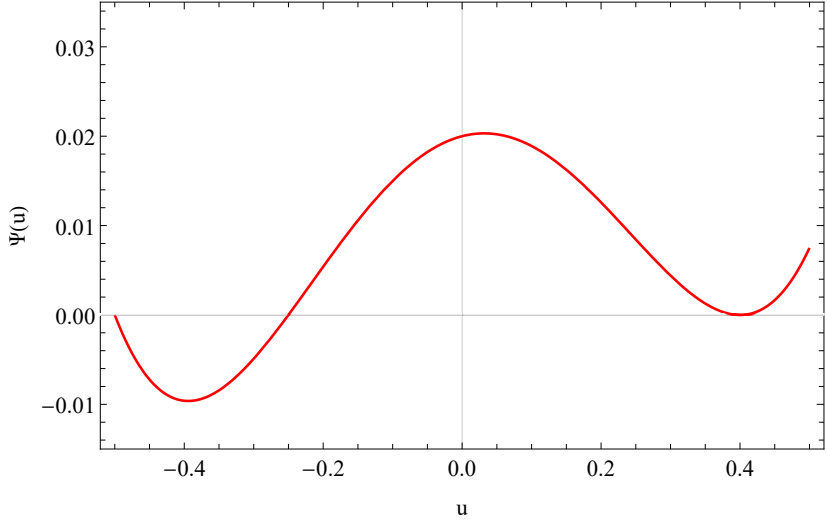}
  \caption{\label{Fig18}\small{\emph{A qualitative illustration of $\Psi(u)$ versus $u$ for QSQ case to see the case of the circular orbits.}}}
\end{figure}

By integrating Eq. (54) with respect to Eq. (58), one can find the trajectory of $u-\phi$ for circular orbits around the QSQ black hole as below
\begin{equation}
\phi=\frac{2}{\sqrt{\left(u_{c}-u_{1}\right)}\sqrt{\left(u_{2}-u_{c}\right)}}\tan^{-1}\left[\frac{\sqrt{\left(u-u_{1}\right)}\sqrt{\left(u_{2}-u_{c}\right)}}
{\sqrt{\left(u-u_{2}\right)}\sqrt{\left(u_{c}-u_{1}\right)}}\right]\,.
\end{equation}

According to Figure 7 and Figure 17, for $E_{n}=E_{2}$, the function of $\Psi(u)$ has four real roots, and for $E_{n}=E_{1}$, it has one real root. These two cases lead to unbounded orbits. For the case $E_{n}=E_{2}$, due to the presence of the black hole, the unbounded null geodesics corresponded to photons path deviate from the original direction so that, they come from infinity to somewhere near the black hole called the closest approach and then deviate to go to some different directions to infinity.

\subsection{Gravitational Lensing as an Application of Null Geodesics}

Now, we focus on the calculation of the closest approach distant for the QSQ black hole. Then, by the obtained result for the closest approach, we compute the corresponding bending angel.

\subsubsection{Closest Approach}

The closest approach distance, noted by $r_{o}$ for a photon, which has the energy $E_{2}$ can be calculated by the relation $\frac{dr}{d\phi}=0$. From Eqs. (42) and (43) in terms of $r$, one can find the following equation in QSQ case
\begin{eqnarray}
\bigg(\frac{1}{r^{2}}\frac{dr}{d\phi}\bigg)^{2}=\Psi(r)=\frac{2M}{r^{3}}-\frac{1}{r^{3}}\sqrt{r^{2}-a^{2}}+\frac{c}{r}+\frac{E_{n}^{2}}{L^{2}}\nonumber\\
=\frac{2M}{r^{3}}-\frac{1}{r^{3}}\sqrt{r^{2}-a^{2}}+\frac{c}{r}+\frac{1}{D^{2}}\,.
\end{eqnarray}
Again, Eq. (60) has not explicit roots. So, in order to eliminate higher-order terms of $a$, one can expand Eq. (60) to produce the following result
\begin{equation}
\left(\frac{1}{r^{2}}\frac{dr}{d\phi}\right)^{2}=\Psi(r)\approx-\frac{1}{r^{2}}+\frac{2M}{r^{3}}+\frac{a^{2}}{2r^{4}}+\frac{c}{r}+\frac{1}{D^{2}}\,,
\end{equation}
Now, the closest approach could be addressed by the roots of the $\Psi(r)$ in Eq. (61). These roots are listed as below
\begin{equation}
r_{1}=-\frac{D^{2}c}{4}-\sqrt{f_{2}}-\frac{1}{12}\sqrt{f_{3}+\frac{9}{2}\frac{\left(D^{6}c^{3}+4D^{4}c+16D^{2}M\right)}{\sqrt{f_{2}}}}\,,
\end{equation}
\begin{equation}
r_{2}=-\frac{D^{2}c}{4}-\sqrt{f_{2}}+\frac{1}{12}\sqrt{f_{3}+\frac{9}{2}\frac{\left(D^{6}c^{3}+4D^{4}c+16D^{2}M\right)}{\sqrt{f_{2}}}}\,,
\end{equation}
\begin{equation}
r_{3}=-\frac{D^{2}c}{4}+\sqrt{f_{2}}-\frac{1}{12}\sqrt{f_{3}-\frac{9}{2}\frac{\left(D^{6}c^{3}+4D^{4}c+16D^{2}M\right)}{\sqrt{f_{2}}}}\,,
\end{equation}
\begin{equation}
r_{4}=-\frac{D^{2}c}{4}+\sqrt{f_{2}}+\frac{1}{12}\sqrt{f_{3}-\frac{9}{2}\frac{\left(D^{6}c^{3}+4D^{4}c+16D^{2}M\right)}{\sqrt{f_{2}}}}\,,
\end{equation}
where we have defined
\begin{eqnarray}
f_{1}\equiv 3\Bigg\{108a^{2}D^{6}c^{2}+288a^{2}D^{4}+\bigg[\Big(108a^{2}D^{6}c^{2}+288a^{2}D^{4}+144D^{6}cM-16D^{6}+864D^{4}M^{2}\Big)^{2}\nonumber\hspace{-1.5 cm}\\-4\Big(24 a^{2}D^{2}-24D^{4}cM+4D^{4}\Big)^3\bigg]^{\frac{1}{2}}+144D^{6}cM-16D^{6}+864D^{4}M^{2}\Bigg\}^{\frac{1}{3}}\,,
\end{eqnarray}
\begin{equation}
f_{2}\equiv\frac{6a^{2}D^{2}-6D^{4}cM+D^{4}}{2^{^{\frac{2}{3}}}f_{1}}+\frac{D^{4}c^{2}}{16}+\frac{D^{2}}{6}+\frac{f_{1}}{72\sqrt[3]{2}}\,,
\end{equation}
\begin{equation}
f_{3}\equiv-\frac{72\sqrt[3]{2}\Big(6a^{2}D^{2}+D^{4}(1-6cM)\Big)}{f_{1}}+18D^{4}c^{2}+48D^{2}-2^{^{\frac{2}{3}}}f_{1}\,.
\end{equation}
Two of these roots, i.e., $r_{1}$, and $r_{2}$ have a negative sign. So, one should eliminate them. We choose the root, $r_{4}$ as the closest approach, $r_{o}$ for the QSQ black hole \cite{Fernando2012}.

\begin{figure}
  \centering
  \includegraphics[width=0.7\textwidth]{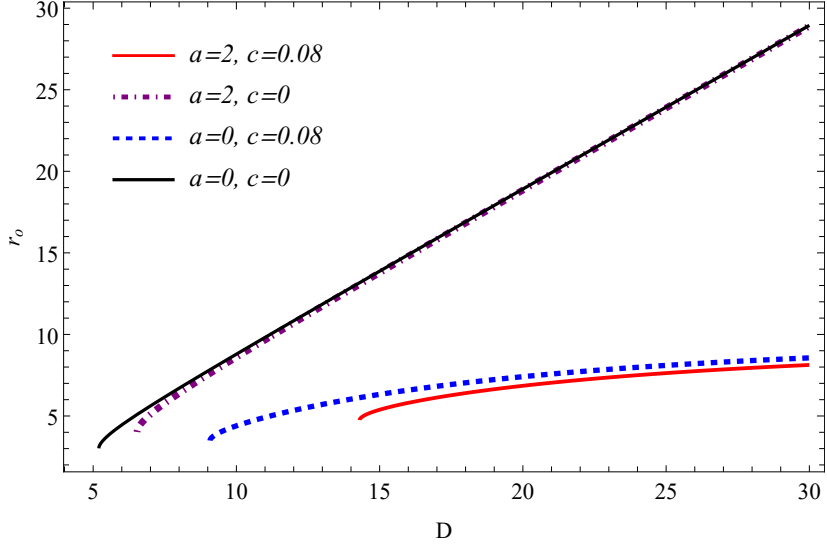}
  \caption{\label{Fig19}\small{\emph{Illustration of $r_{o}$ versus $D$ for QSQ, QS, SQ, and SCH cases in which we set $M=1$.}}}
\end{figure}
Figure 19 shows the graph of $r_{o}$ versus impact parameter, $D$ for QSQ, QS, SQ, and SCH cases. The values of $r_{o}$ in the QSQ case are the same as the SQ case, approximately. But, the values of $r_{o}$ in the QSQ case begin from some bigger values of $D$. Moreover, it seems that the role of the quintessential matter is to cut off the amounts of $r_{o}$, while for the SCH and QS cases, $r_{o}$ goes to infinite, continually.

\subsubsection{Bending Angle}

Consider a light ray, which starts from a remote point (e.g., an actual position of a star, a supernova and so on) than the black hole in an asymptotically flat region of spacetime, and then approaches the black hole. At the closest approach distance, $r_{o}$, the light ray becomes deflected by its strong attraction to reach to a far observer in another point (in an asymptotically flat region). The angle of the deflection is known as the bending angle, $\alpha$. Figure 20 depicts such an angle, which is responsible for the apparent location of that remote point for the far observer.
\begin{figure}
  \centering
  \includegraphics[width=0.8\textwidth]{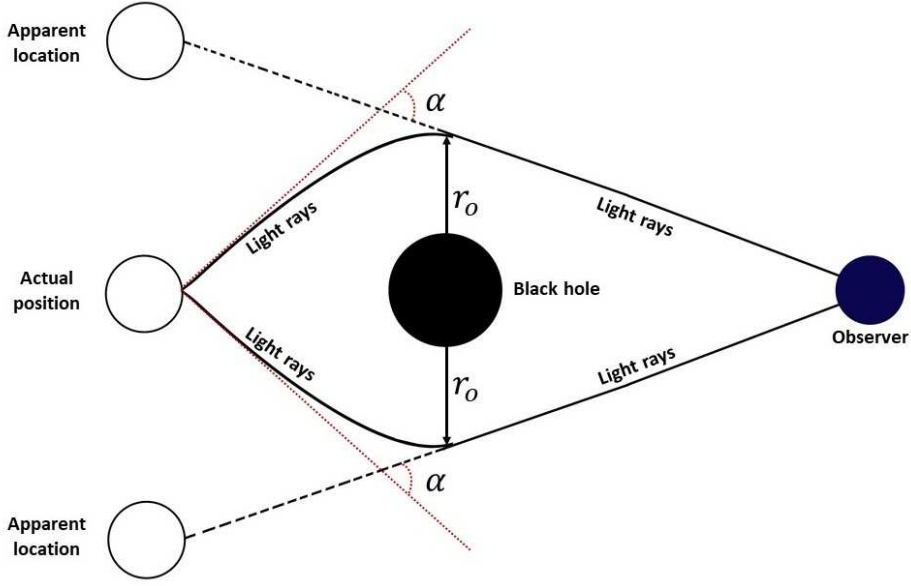}
  \caption{\label{Fig20}\small{\emph{The illustration of the bending angle, $\alpha$.}}}
\end{figure}
From Eq. (18), one can write the bending angle, $\alpha$ for the QSQ black hole in terms of the expanded $\Psi(r)$ in Eq. (61) as follows \cite{Keeton2005}
\begin{eqnarray}
\alpha=2\int_{r_{o}}^{\infty}\bigg|\frac{d\phi}{dr}\bigg|dr-\pi\nonumber\hspace{7.7 cm}\\
=2\int_{r_{o}}^{\infty}\frac{1}{r^{2}}\frac{dr}{\sqrt{\big({E_{n}}/{L^{2}}\big)^{2}-\big(g_{_{qq}}(r)/{r^{2}}\big)}}-\pi=
2\int_{r_{o}}^{\infty}\frac{1}{r^{2}}\frac{dr}{\sqrt{\Psi(r)}}-\pi\,.
\end{eqnarray}
By using Eqs. (44) and (52), one can rewrite Eq. (69) in terms of variable $u$ as follows \cite{Azreg2017}
\begin{equation}
\alpha=2\int_{0}^{u_{o}}\frac{du}{\sqrt{\Psi(u)}}-\pi\,,
\end{equation}
where $u_{o}=\frac{1}{r_{o}}$. This integral easily could be done as follows
\begin{equation}
\alpha=2\frac{\mathrm{EllipticK}[w]}{\sqrt{\left(u_{1}-u_{3}\right)}\sqrt{\left(u_{2}-u_{o}\right)}}-2\frac{\mathrm{EllipticF}[\rho,y]}{\sqrt{\left(u_{1}-u_{3}\right)}\sqrt{\left(u_{2}-u_{o}\right)}}-\pi\,,
\end{equation}
in which we have defined
\begin{equation}
w\equiv\frac{\left(u_{1}-u_{o}\right)\left(u_{2}-u_{3}\right)}{\left(u_{1}-u_{3}\right)\left(u_{2}-u_{o}\right)}\,,
\end{equation}
\begin{equation}
\sin\rho\equiv\frac{\sqrt{u_{1}\left(u_{2}-u_{o}\right)}}{\sqrt{u_{2}\left(u_{1}-u_{o}\right)}}\,,
\end{equation}
\begin{equation}
y\equiv\frac{\left(u_{1}-u_{o}\right)\left(u_{2}-u_{3}\right)}{\left(u_{1}-u_{3}\right)\left(u_{2}-u_{o}\right)}\,,
\end{equation}
and $\mathrm{EllipticK}[w]$ is the Complete Elliptic Integral of the first kind, and $\mathrm{EllipticF}[\rho,y]$ is the Elliptic Integral of the first kind. By numerical analyses for the Sun, which have a bending angle about $\alpha=1.75\,\,\mathrm{arcsec}$, and by setting $D=1$ and $c=0.001$, one can find a numerical range for the quantum correction parameter, $a$ as $$10^{-10}<a<10^{-1}\,,$$ approximately. However, in the present study, we choose $a=2$ to have better illustrations for different quantities.

\section{Time-Like Geodesics Structure}

In this section, we focus on the time-like geodesics for the QSQ black hole to analyze the motion of a massive test particle. By using Eqs. (10) and (11) with respect to the QSQ metric coefficient (4), one can find the geodesics equations for the time-like case in this scheme as follows \cite{Uniyal2015}
\begin{equation}
\ddot{t}+\frac{g'(r)}{g(r)}\dot{t}\dot{r}=0\,,
\end{equation}
\begin{equation}
\ddot{r}+\frac{1}{2}g(r)g'(r)\dot{t}^{2}-\frac{1}{2g(r)}g'(r)\dot{r}^{2}-g(r)r\dot{\theta}^{2}-g(r)r\sin^{2}\theta\dot{\phi}^{2}=0\,,
\end{equation}
\begin{equation}
\ddot{\theta}+\frac{2}{r}\dot{r}\dot{\theta}-\sin\theta\,\cos\theta\,\dot{\varphi}^{2}=0\,,
\end{equation}
\begin{equation}
\ddot{\varphi}+\frac{2}{r}\dot{r}\dot{\phi}+2\cot\theta\,\dot{\theta}\dot{\varphi}=0\,,
\end{equation}
where `prime' denotes the differentiation with respect to the radial coordinate, $r$. From Eqs. (3), (4), and (11), one can write the constraint equation for the time-like geodesics with $e=-1$ as follows
\begin{equation}
-g(r)\left(\frac{dt}{ds}\right)^{2}+\frac{1}{g(r)}\left(\frac{dr}{ds}\right)^{2}+r^{2}\left(\frac{d\theta}{ds}
\right)^{2}+r^{2}\sin^{2}(\theta)\left(\frac{d\phi}{ds}\right)^{2}=-1\,.
\end{equation}
Now, we study the time-like geodesics equations on the equatorial plane.

\subsection{Time-Like Geodesics Equations on the Equatorial Plane}

We set $\theta=\frac{\pi}{2}$ to analyze the time-like geodesics equations on the equatorial plane for the QSQ black hole by integrating Eqs. (75) and (78), which results in the following equations
\begin{equation}
\dot{t}=\frac{Z_{1}}{\left(-\frac{2M}{r}+\frac{1}{r}\sqrt{r^{2}-a^{2}}-cr\right)}\,,
\end{equation}
\begin{equation}
\dot{\varphi}=\frac{Z_{2}}{r^{2}}\,,
\end{equation}
where $Z_{1}$ and $Z_{2}$ are the constants of integrations, which correspond to two conserved quantity: the total energy $E$, and the total angular momentum $L$ of the test particle traveling on the time-like geodesics, respectively \cite{Uniyal2015}. By substituting Eqs. (80) and (81) in the constraint equation (79), one can obtain the equation of energy conservation as follows \cite{Uniyal2015}
\begin{equation}
E^{2}=\dot{r}^{2}+V_{eff}(r)\,,
\end{equation}
where $V_{eff}(r)$ is defined as follows
\begin{eqnarray}
V_{eff}(r)=g(r)\bigg(\frac{L^{2}}{r^{2}}+1\bigg)=\bigg(-\frac{2M}{r}+\frac{1}{r}\sqrt{r^{2}-a^{2}}-c\bigg)\bigg(\frac{L^{2}}{r^{2}}+1\bigg)
\nonumber\hspace{-0.5 cm}\\=-\frac{2M}{r}-\frac{2ML^{2}}{r^{3}}+\frac{1}{r}\sqrt{r^{2}-a^{2}}\bigg(\frac{L^{2}}{r^{2}}+1\bigg)-c\bigg(\frac{L^{2}}{r^{2}}+1\bigg)\,.
\end{eqnarray}
Henceforward, we focus on the study of the time-like geodesics with angular momentum (i.e., $L\neq 0$).

\subsubsection{Effective Potential, $V_{eff}$}

Figure 21 shows that the curve of $V_{eff}$ versus $r$ in the QSQ case is smaller than the SQ case. Also, it seems that due to the quantum deformation term in the scheme, the curves of QSQ and QS cases shifted to $r>a$, where we set $a=2$. So again, it seems that the reduction, and also, shifting of the $V_{eff}$ function is due to the presence of quantum effects in this scenario.
\begin{figure}
  \centering
  \includegraphics[width=0.7\textwidth]{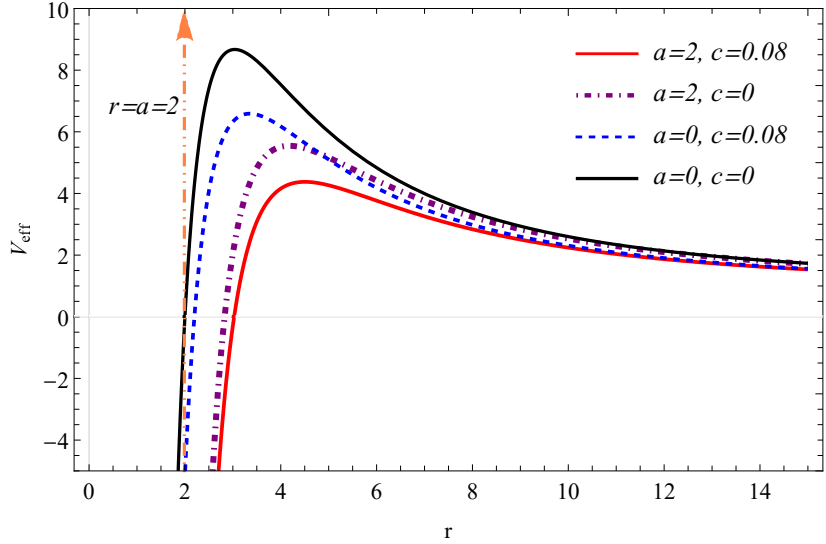}
  \caption{\label{Fig21}\small{\emph{The illustration of $V_{eff}$ versus $r$ for QSQ, QS, SQ, and SCH cases in which we set $L=15$ and $M=1$.}}}
\end{figure}

\subsubsection{Unstable, Circular Orbits}

Due to the nature of the effective potential, from Figure 22, one can see that the motion of the test particle is not stable for the time-like geodesics at $r_{c}$, and also, from Eq. (82), we have $E^{2}-V_{eff}(r)=\dot{r}^{2}=0$.
\begin{figure}
  \centering
  \includegraphics[width=0.7\textwidth]{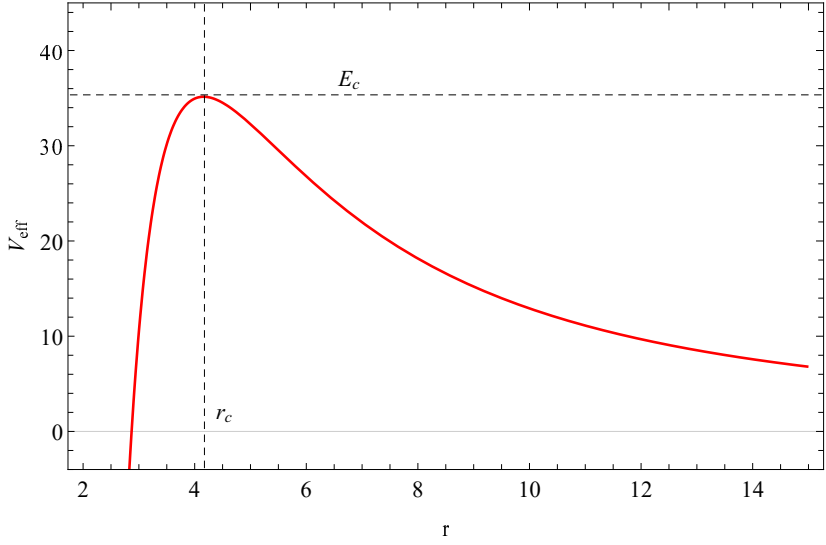}
  \caption{\label{Fig22}\small{\emph{The illustration of $V_{eff}$ versus $r$ for QSQ case in which we set $a=2$, $c=0.02$, $L=40$ and $M=1$ to see where the unstable, circular orbits occur.}}}
\end{figure}

So, $r_{c}$ is the unstable, circular orbit for the time-like geodesics around QSQ black hole. At this point, from Eq. (25), we have
\begin{equation}
\frac{dV_{eff}}{dr}=0\,.
\end{equation}
By solving Eq. (84), one can find its solutions for $r$ to read the $r_{c}$. But Eq. (84) with respect to Eq. (83) has no explicit solution. Hence, we expand Eq. (83) to eliminate higher-order terms of $a$ as follows
\begin{equation}
V_{eff}=-\frac{2M}{r}-\frac{2ML^{2}}{r^{3}}+\frac{L^{2}+r^{2}}{r^{2}}-\frac{a^{2}\left(L^2+r^{2}\right)}{2r^{4}}-c\left(\frac{L^{2}}{r^{2}}+1\right)\,.
\end{equation}

Now, one can solve Eq. (84) in terms of Eq. (85) to find its solutions as are listed below
\begin{equation}
r_{1}=\frac{W_{1}}{6\sqrt[3]{2}M}-\frac{W_{2}}{6M}+W_{3}\,,
\end{equation}
\begin{equation}
r_{2}=-\left(1-i\sqrt{3}\right)\frac{W_{1}}{12\sqrt[3]{2}M}-\frac{W_{2}}{6M}-\left(1+i\sqrt{3}\right)\frac{W_{3}}{2}\,,
\end{equation}
\begin{equation}
r_{3}=-\left(1+i\sqrt{3}\right)\frac{W_{1}}{12\sqrt[3]{2}M}-\frac{W_{2}}{6M}-\left(1-i\sqrt{3}\right)\frac{W_{3}}{2}\,,
\end{equation}
where we have defined
\begin{eqnarray}
W_{1}\equiv\sqrt[3]{2}\Bigg\{-a^{6}-6a^{4}(c-1)L^{2}-6a^{2}\Big(2(c-1)^{2}L^{4}+9L^{2}M^{2}\Big)-8(c-1)^{3}L^{6}\nonumber\hspace{-1 cm}\\+108(c-1)L^{4}M^{2}+6\sqrt{3}\,LM\bigg[2a^{8}+12a^{6}(c-1)L^{2}+3a^{4}\big(8(c-1)^{2}L^{4}-3L^{2}M^{2}\big)
\nonumber\hspace{-1 cm}\\+4a^{2}(c-1)L^{4}\Big(4(c-1)^{2}L^{2}-63M^{2}\Big)-36L^{4}M^{2}\Big((c-1)^{2}L^{2}
-12M^{2}\Big)\bigg]^{\frac{1}{2}}\Bigg\}^{\frac{1}{3}}\,,
\end{eqnarray}
\begin{equation}
W_{2}\equiv a^{2}+2cL^{2}-2L^{2}\,,
\end{equation}
\begin{equation}
W_{3}\equiv\frac{W_{2}^{2}-36L^{2}M^{2}}{3M2^{^{2/3}}W_{1}}\,.
\end{equation}
The sign of $r_{2}$ is negative, and we choose $r_{3}$ as the radius of unstable, time-like circular orbits, $r_{c}$.

Figure 23 depicts the curve of the $r_{c}$ in terms of $c$ for QSQ and SQ cases. It seems that the QSQ curve in comparison to the SQ case has lager values for $r_{c}$.
\begin{figure}
  \centering
  \includegraphics[width=0.7\textwidth]{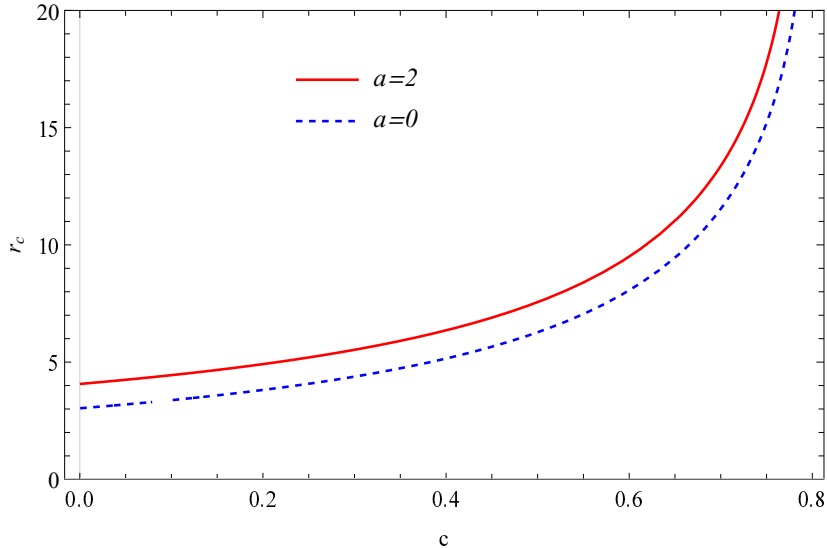}
  \caption{\label{Fig23}\small{\emph{The illustration of $r_{c}$ versus $c$ for QSQ and SQ cases in which we set $L=17$ and $M=1$.}}}
\end{figure}

\subsection{Force on the Test Particle}

From Eq. (83), one can find the force on the test particle for QSQ black hole as follows
\begin{eqnarray}
F_{eff}=-\frac{1}{2}\frac{dV_{eff}}{dr}\nonumber\hspace{11.09 cm}\\=
-\frac{M}{r^{2}}\bigg(1+\frac{3L^{2}}{r^{2}}\bigg)-\frac{cL^{2}}{r^{3}}+\frac{2L^{2}\big(r^{2}-a^{2}\big)+
\big(r^{2}+L^{2}\big)\big(r^{2}-a^{2}\big)-\big(r^{4}+r^{2}L^{2}\big)}{2r^{4}\sqrt{\big(r^{2}-a^{2}\big)}}\,,
\end{eqnarray}
where from the equation of motion (82), we have inserted a factor of $\big(\frac{1}{2}\big)$ \cite{Fernando2012}. In Eq. (92), the first term is the Newtonian term. The sign of the second term is negative, hence it is attractive. This term is associated with the quintessential matter, which is a candidate for dark energy. So, the dark energy component of the Universe in this study is attractive. The last term in Eq. (95) associated with the quantum effects is a repulsive force, due to its positive sign, as the previous section.

Figure 24 is the plot of $F_{eff}$ versus $r$ for QSQ, QS, SQ, and SCH cases. It is seen that the curves of QSQ and QS cases are shifted to the $r=a=2$, due to the presence of the quantum effects in this study, and also, they are smaller than the other cases.
\begin{figure}
  \centering
  \includegraphics[width=0.7\textwidth]{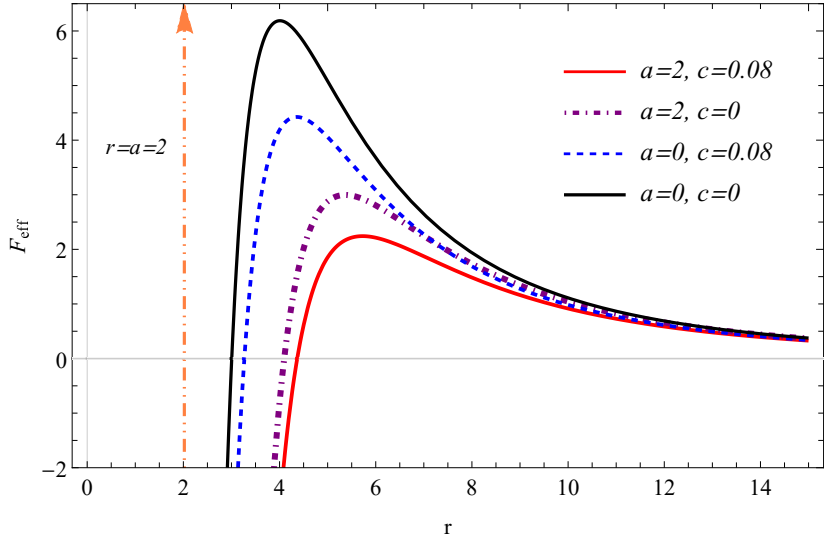}
  \caption{\label{Fig24}\small{\emph{The illustration of $F_{eff}$ versus $r$ for QSQ, QS, SQ, and SCH cases in which we set $L=40$ and $M=1$.}}}
\end{figure}

\section{Summary and Conclusions}

It is well known that adding quantum effects into the framework of the general theory of relativity leads to solve some controversial puzzles within this theory. Accordingly, by taking into account the quantum excitations of the background metric and matter fields, one can remove the undesirable spacetime singularity at the center of the SCH black hole. In this manner, Kazakov and Solodukhin found the QS black hole in which the singularity point converts to a two-dimensional sphere of radius $r\sim l_{Pl}$. On the other hand, due to the increasing acceleration of the expansion of the Universe, it seems that there exists a repulsive force known as dark energy. Kiselev considered the quintessence field, as a candidate model for describing dark energy, in the background of the SCH black hole. Based on these two ideas, Shahjalal took into account the QSQ black hole to investigate its thermodynamics. On the other hand, investigating the geodesic structure of black holes is a way to understand, and test the physical effects and behavior in the spacetime. In this study, we have considered the QSQ black hole to study the corresponding null and time-like geodesics structures, and then we compared the achieved results to the outcomes of the Refs. \cite{Fernando2012, Uniyal2015}.

We have studied the null, circular orbits, and also, we have computed the Lyapunov exponent for the QSQ black hole. It is shown that the presence of quintessence matter leads to make the cosmological horizon. Also, the spacetime becomes static among the event horizon and the cosmological horizon, where it is interesting to study the motion of particles. We have shown that the quantum effects in the scheme result in shifting the inner structure of the SCH black hole by converting its central point-like singularity to a two-dimensional sphere of radius $r=a\sim l_{Pl}$. Consequently, the world line of the massless and massive particles are shifted to farther from $r=0$ than the non-deformed cases. We also have shown that based on the correspondence principle, the QS case tends to the SCH one, and the QSQ case approaches the SQ case, respectively. We have proved that because of the quantum fluctuations of the background metric, the period of null, circular orbits in the QSQ and the QS cases are always non-zero even in the absence of the black hole. We also have proved that quantum effects in this setup are responsible for accelerating the expansion of the Universe, while the quintessence field results in decelerating the expansion. So, quantum effects, together with the quintessence field result in regulating and limiting the acceleration of the expansion of the Universe. Also, it seems that due to the presence of quantum effects and quintessence in this scheme, the values of Hawking temperature, effective potential, and also, effective force have been reduced significantly. By numerical analysis, we have shown that the presence of cosmological horizon leads to value the parameter of quantum corrections as $a\approx 10^{-9}$, and also, the gravitational lensing in the case of Sun results in the range of $10^{-10}<a<10^{-1}$ for this parameter.\\

{\bf Acknowledgement}\\
The authors would like to thank Akram Sadat Sefiedgar and Sara Saghafi for fruitful discussions. The authors also would like to thank Raziyeh Safiyan for drawing Figure 20.

\end{document}